\documentclass[journal=jacsat,manuscript=article]{achemso}
\usepackage{graphicx} 
\usepackage{chemformula} 
\usepackage[T1]{fontenc} 

\SectionNumbersOn 
\usepackage[normalem]{ulem}
\usepackage{multirow,booktabs,tabularray}
\usepackage{adjustbox}
\usepackage{colortbl}
\usepackage{array}
\usepackage{subfigure}
\usepackage{amssymb}
\usepackage{pdfpages}

\setkeys{acs}{
etalmode = truncate,   
maxauthors = 10        
}
\newcommand{\mz}[1]{{\itshape m/z}\,\textup{#1}}

\title[An \textsf{achemso} demo]
{Acenaphthene Derivatives as Signatures of \ch{C11H9+} Reactivity with Methylated Naphthalenes}

\author{Ana I. Lozano}
\author{Anthony Bonnamy}
\affiliation[Universit\'e de Toulouse]
{Institut de Recherche en Astrophysique et Plan\'etologie (IRAP), Universit\'e de Toulouse, CNRS, CNES, 9 Avenue du Colonel Roche, 31028 Toulouse, France}
\author{Aude Simon}
\affiliation[Universit\'e de Toulouse]
{Laboratoire de Chimie et Physique Quantiques LCPQ/FeRMI, Universit\'e de Toulouse, CNRS, 118 Route de Narbonne, 31062 Toulouse, France}
\author{Christine Joblin}
\email{christine.joblin@cnrs.fr}
\affiliation[Universit\'e de Toulouse]
{Institut de Recherche en Astrophysique et Plan\'etologie (IRAP), Universit\'e de Toulouse, CNRS, CNES, 9 Avenue du Colonel Roche, 31028 Toulouse, France}

\begin{document}

\begin{abstract}

\ch{C11H9+} ion is the dominant fragment cation formed from methyl-naphthalene (MeNp) and dimethyl-naphthalene (diMeNp).
Using the multiplex capabilities of PIRENEA, a setup dedicated to laboratory astrophysics, we studied the reactivity of the benzylium-like isomers of \ch{C11H9+} with diMeNp under isolated conditions relevant to radiative association. Two reaction products are observed, \ch{C12H11+} -- also formed in the reaction with MeNp -- and \ch{C13H13+}, with branching ratios that depend on the specific diMeNp isomer.
The reaction products were subsequently exposed to UV–visible irradiation to gain insight into their structures. The acenaphthylene radical cation, \ch{C12H8^{.}+}, was identified as the most stable photofragment. We show that this experimental approach, supported by density functional theory calculations and molecular dynamics simulations, provides new constraints on the chemistry of benzylium-type species. We highlight the role that long-lived ion–molecule complexes can have in promoting C–C coupling and the formation of a pentagonal cycle.
Moreover, the chemistry uncovered here highlights new pathways for the formation of pentagonal rings during PAH growth under low-pressure and cold conditions. In particular, it can lead to efficient formation of acenaphthylene-like species, recently detected in the TMC-1 cold cloud.
\end{abstract}

Keywords: ion-molecule reactions, long-lived complexes, photofragments, molecular dynamics simulations, polycyclic aromatic hydrocarbons, pentagonal rings, benzylium/tropylium, astrochemistry.

\section{Introduction}

A recent milestone in astrochemistry is the identification of individual gas-phase polycyclic aromatic hydrocarbons (PAHs) in the cold, dense cloud TMC-1. Using the Green Bank Telescope (GBT), McGuire et al. reported the detection of 1- and 2-cyanonaphthalene\cite{McGuire2021}. Shortly afterward, the pure bicyclic aromatic indene was detected by Cernicharo et al.\cite{Cernicharo2021} with the highly sensitive receivers of the Yebes 40 m radio telescope, and this detection was soon confirmed at the GBT\cite{Burkhart2021}. Since then, additional PAHs have been identified, including cyano-acenaphthylene \ch{C12H7CN} (1- and 5- isomers)\cite{Cernicharo2024}, phenalene \ch{C13H10}\cite{Cabezas2025}, cyano-pyrene \ch{C16H9CN} (1-, 2- and 4- isomers)\cite{Wenzel2024, Wenzel2025} and cyano-coronene \ch{C24H11CN}, the largest PAH identified so far\cite{Wenzel2025ApJ}. Most PAHs detected to date are cyano-substituted species, whose large dipole moments significantly enhance their radio detectability.

The formation of cyano-PAHs under the cold conditions of dark molecular clouds is thought to result from reactions between PAHs and CN radicals; for benzene and toluene, this reaction has been shown to be barrierless\cite{Balucani1999,Cooke2020,Messinger2020}. This connection makes it possible to infer the abundances of the unsubstituted parent PAHs from the observed abundances of their cyano derivatives\cite{Wenzel2025, Wenzel2025ApJ}. The inferred values indicate that PAHs must be produced efficiently in cold clouds\cite{Agundez2023}. Although the underlying processes remain uncertain, current evidence favors a bottom-up gas-phase chemistry\cite{Cernicharo2022,Cernicharo2024,Cabezas2025}. Moreover, consistent with the pioneering work of Herbst and Klemperer on dense-cloud chemistry\cite{Herbst1973}, these findings further highlight the key role of ion–molecule reactions in building molecular complexity\cite{Mallo2025}.

PAHs -- including methylated and, more broadly, alkylated PAHs -- have been found to be abundant in laboratory analyses of primitive Solar System matter, particularly in carbonaceous chondrite meteorites\cite{Naraoka2000,Plows2003,Elsila2005,Danger2020,Lecasble2022} and, more recently, in samples from the asteroid Ryugu returned by the Hayabusa-2 mission\cite{Aponte2023, Sabbah2024}. Zeichner et al.\cite{Zeichner2023} investigated the $^{13}$C isotopic composition of several small PAHs in Ryugu material and concluded that these molecules were formed through distinct chemical processes: some at the very low temperatures characteristic of molecular clouds ($\sim$10~K), and others at much higher temperatures (>1000~K) that may prevail in circumstellar environments or within the parent body.

Methylated PAHs may contribute to the 3.4~$\mu$m emission band\cite{Joblin1996}, which appears as a satellite feature of the 3.3~$\mu$m aromatic infrared band (AIB) -- both of which are key targets of James Webb Space Telescope (JWST) observations in the Milky Way and in external galaxies\cite{Peeters2024,Schroetter2024,Thatte2025}. The 3.4~$\mu$m band is known to be particularly sensitive to UV irradiation\cite{Joblin1996, Pilleri2015}, consistent with its carriers containing aliphatic C–H bonds, either in alkyl side groups or in the form of superhydrogenated PAHs with additional H atoms\cite{Jochims1999,Marciniak2021}.

Previously, we investigated the properties of the \ch{C11H9+} ions -- the H-loss fragments of methyl-naphthalene cations -- and highlighted the relevance of these ions in environments exposed to mild UV irradiation due to their photophysical properties\cite{lozano2025}. Three long-lived isomers were identified: 1- and 2-naphthylmethylium (1- and 2-\ch{NpCH2+}) and benzyltropylium (\ch{BzTr+}). Their reactions with the parent neutral lead to two products of unknown structure, \ch{C12H11+} and \ch{C22H19+}, the latter corresponding to an adduct. It was further shown that \ch{BzTr+} is unreactive, whereas the other two isomeric populations -- very likely 1- and 2-\ch{NpCH2+} -- display reaction rate constants differing by an order of magnitude.

This article investigates the reactivity of \ch{C11H9+} ions with several isomers of dimethyl-naphthalene (diMeNp), namely 2,3-, 1,4-, and 1,5-diMeNp. Using the multiplex capabilities of PIRENEA, a setup dedicated to laboratory astrophysics, we studied these reactions under isolated conditions relevant to radiative association. Moreover, the reaction products were exposed to UV–visible light to gain further insight into their structures and to identify fragments with enhanced photostability. We show that this experimental approach, supported by DFT calculations, not only improves our understanding of the chemistry of these species -- including isomer-specific effects -- but also reveals new chemical pathways for the formation of cosmic PAHs under cold and mild-UV irradiation conditions.

\begin{figure}[ht]
    \centering
   \includegraphics[width=0.7\linewidth]{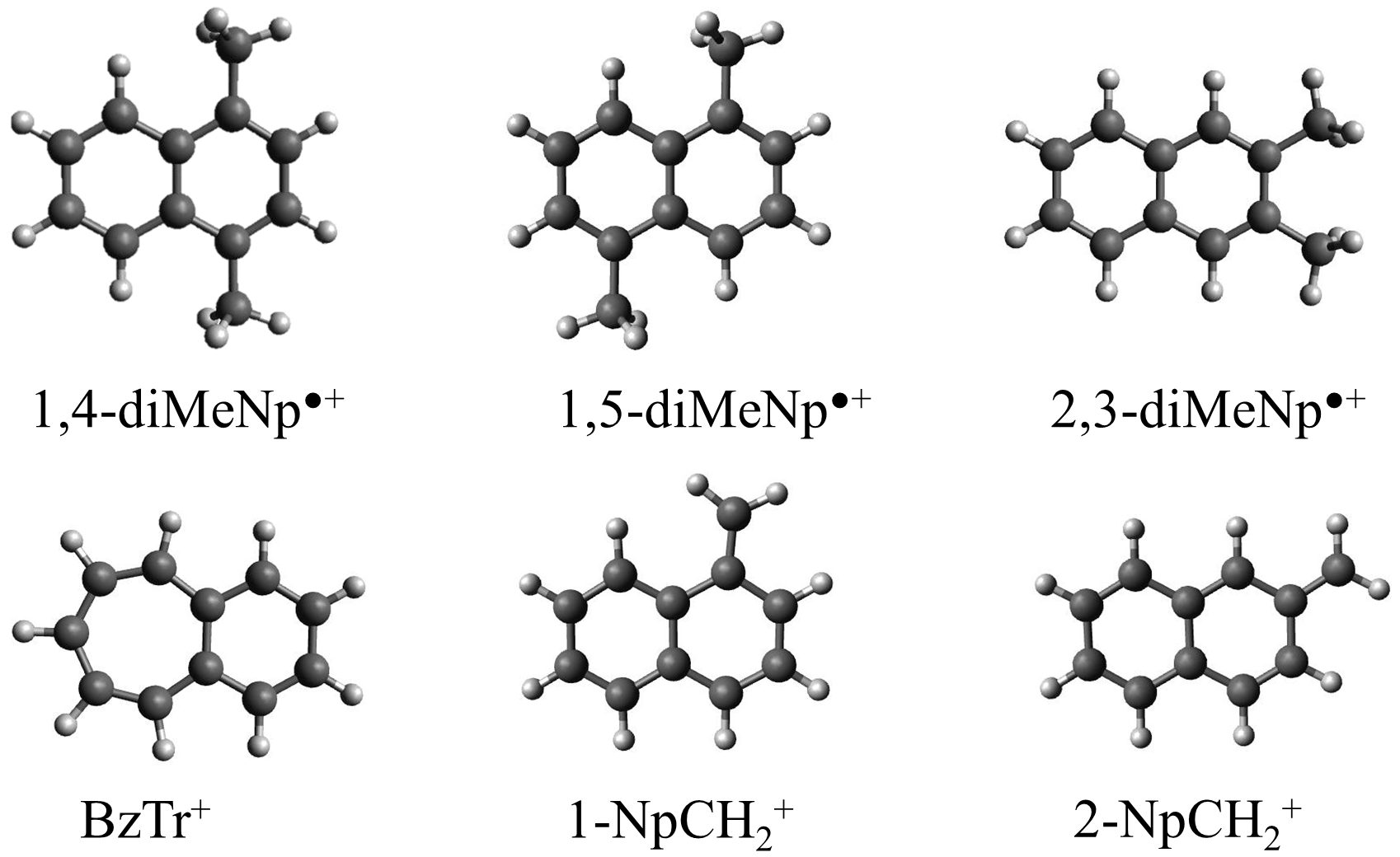}
    \caption{Top row: molecular structures of 1,4-, 1,5-, and 2,3-dimethylnaphthalene cations (1,4-, 1,5, and 2,3-\ch{diMeNp^{.}+}). Bottom row: molecular structures of the three expected long-lived isomers of \ch{C11H9+} cations: benzyltropylium (\ch{BzTr+}), 1-, and 2-naphthylmethylium (1- and 2-\ch{NpCH2+}). }
    \label{molecules}
\end{figure}

\section{Methodologies}

\subsection{Experimental Methods}
\label{Experimental}

The experiments were performed using the PIRENEA setup\cite{Useli-Bacchitta2010}, operated under the configuration described in detail elsewhere\cite{lozano2025}. 
The apparatus consists of a Fourier Transform Ion Cyclotron Resonance Mass Spectrometer (FTICR-MS) maintained under ultra-high vacuum conditions ($\sim 10^{-9}$ mbar at room temperature), enabling ion trapping under isolated conditions and for long time scales. PIRENEA can also be run at cryogenic temperatures but this capability was not used in this study due to the sticky nature of the injected molecules.
When required, specific ions in the ICR cell (e.g., $^{12}$C isotopologs) were isolated by selectively exciting the cyclotron motion of unwanted ions, thereby ejecting them.

\ch{C11H9+} cations were produced from 2,3-, 1,4-, and 1,5-diMeNp (Figure~\ref{molecules}), all of which possess sufficient vapor pressure at room temperature to be used in our experiments.
The neutral precursors were ionized using the fourth harmonic (266~nm, 4.66~eV) of a Nd:YAG Surelite I laser (Continuum), producing ions that were trapped in the ICR cell. The pure $^{12}$C parent ions were then isolated, and the production of \ch{^{12}C11H9+} was optimized through photoprocessing of the trapped parent ions using a high-pressure Xe arc lamp equipped with a 610~nm color filter (CF). Under these conditions, the parent ions dissociate at low energies, and the resulting fragment ions do not absorb further. Consequently, the \ch{^{12}C11H9+} fragments are expected to have low internal energy and thermalize with the background during ion production, which occurs over 20\,s. Typical mass spectra illustrating the different steps in the case of 1,4-diMeNp are shown in Figure S1. 

Reactions between \ch{^{12}C11H9+} ions and their neutral precursors were conducted at room temperature and a typical pressure of $\sim 10^{-8}$~mbar. The reaction time was adjusted between 30 and 120\,s to ensure complete conversion of the reactive ion population. Sufficient time (up to 200\,s) was allowed to recover a low background pressure (below a few 10$^{-9}$ mbar) before recording the mass spectrum.
 
Photodissociation kinetics were measured for \ch{^{12}C11H9+} and its reaction cationic products using broadband visible irradiation from a Xe arc lamp. A 400~nm CF filter was used, with or without a neutral-density filter (NDF) providing 37\% transmission.
 
Finally, our most challenging experiment imvolved recording a multiple-photon dissociation (MPD) spectrum of one the reactivity product, \ch{C13H13+}. After the reaction and sufficient pumping time, the product of interest was isolated and irradiated using a mid-band tunable optical parametric oscillator (OPO) laser (Panther EX, Continuum), operating at 10~Hz with 5~ns pulse duration. The dissociation yield was recorded as a function of OPO wavelength using a fixed irradiation time of 10 s (100 laser pulses) and a pulse energy of 5.5~mJ. This multi-stage mass spectrometry (MS$^n$) procedure required 3~min per recorded wavelength.

For all experiments, the recorded intensities were normalized to the summed ion intensity to account for variations in absolute ion intensities across different measurements.

\subsection{Computational Methods}
\label{Theory}

Following the methodology proposed by \citet{Simon2017}, we performed molecular dynamics (MD) simulations to investigate the dissociation of the three diMeNp cations studied here.
In this approach, molecular dissociation is modeled in the electronic ground state, with the internal energy represented as vibrational excitation.
This regime is consistent with the building of internal energy in the cations by absorption of multiple photons from the Xe lamp\cite{Boissel1997}.
However, the energies need to be sufficiently high for dissociation to occur within the nanosecond timescale accessible to the simulation.
This approach has been recently applied to rationalize the dissociation dynamics of both MeNp radical cations (\ch{C11H10^{.}+}) and their \ch{C11H9+} fragments\cite{lozano2025}.

Briefly, the MD simulations were carried out in the Born–Oppenheimer approximation;  the electronic structure is computed on-the-fly using the self-consistent charge density functional based tight-binding (SCC-DFTB) method\cite{Elstner1998}, employing the recently revisedC–H parameter set \cite{Joblin2020}. A Fermi temperature of 1500\,K was applied to prevent oscillations in the self-consistent DFTB procedure, which frequently occur in  dissociated or near-dissociated systems. This also ensured continuity of energies and gradients during level crossing. \cite{Simon2017}
This explicit treatment of the electronic structure enables the description of chemical evolution processes such as isomerization and dissociation.
For simplicity, this approach will hereafter be referred to as MD/DFTB.

Each simulation was initiated from the ground-state optimized geometry with randomized initial velocities, propagated in the microcanonical (NVE) ensemble. 
Energies and gradients were computed every 0.2~fs. 
All MD/DFTB simulations were performed using the deMonNano code~\cite{deMon}.
In the present work, a total of 180 simulations were run for the three diMeNp cations at internal energies of 15.0 and 16.5~eV, for simulation durations of 2~ns and 1~ns, respectively. 
This strategy enables insight into reaction kinetics, branching ratios, and underlying mechanisms and thus contributes to the rationalization of experimental results.

In addition to MD/DFTB simulations, DFT local optimisations and all atoms harmonic frequency calculations were performed at the B3LYP\cite{Becke1993}/6-31G(d,p) level of theory to support and interpret experimental results. Unless it is explicitly mentioned, all reported molecular structures were obtained at this level of theory and all reported energy values include zero-point energy corrections. In specific cases,  wavefunction (CCSD(T)/cc-pvtz) single point energy calculations were performed. Vertical excitation energies were computed using the Time-Dependent (TD) DFT method with the PBE0 functional \cite{Adamo1999} in conjunction with the cc-pvtz basis set. This level of theory has been shown to yield satisfactory excitation energies for \ch{C11H9+} isomers (see Table~S3 in the Supplementary material of \citet{lozano2025}).  All these calculations were achieved with the Gaussian16 suite of programs\cite{g16}.  

\section{Results}
\label{results}

We present our results on the production of \ch{C11H9+} species from the various investigated dimethylnaphthalene ions (Figure~\ref{molecules}), providing evidence for the formation of multiple isomers. These observations are interpreted in the context of previous experimental studies and MD simulations. We then examine the reactions between \ch{C11H9+} and their neutral diMeNp precursors, followed by a detailed characterization of the photophysics of the resulting products.

\subsection {\ch{C11H9+} as a common photofragment}
\label{results141}

\subsubsection{Experimental Insights}
\label{results141_exp}

Our first key experimental finding is that photodissociation of the diMeNp cations yields \ch{C11H9+} at {\it m/z} 141 as the dominant fragment, regardless of the methyl substitution pattern. A similar behavior is observed for the MeNp cations, for which \ch{C11H9+} is produced as the exclusive photofragment\cite{Gotkis1993, lozano2025}.
In all cases, the \ch{C11H9+} cation further dissociates by losing \ch{C2H2}, producing \ch{C9H7+} cations, which were previously assigned to indenylium when 1-MeNp was used as the precursor\cite{lozano2025}.
Unlike the photodissociation of MeNp cations, a very minor H-loss channel was detected for diMeNp when irradiated at 266~nm (see  Figure S1), accounting for at most $\sim$3\% of the total dissociation yield.

\begin{figure}[!ht]
    \centering
   \includegraphics[width=0.6\linewidth]{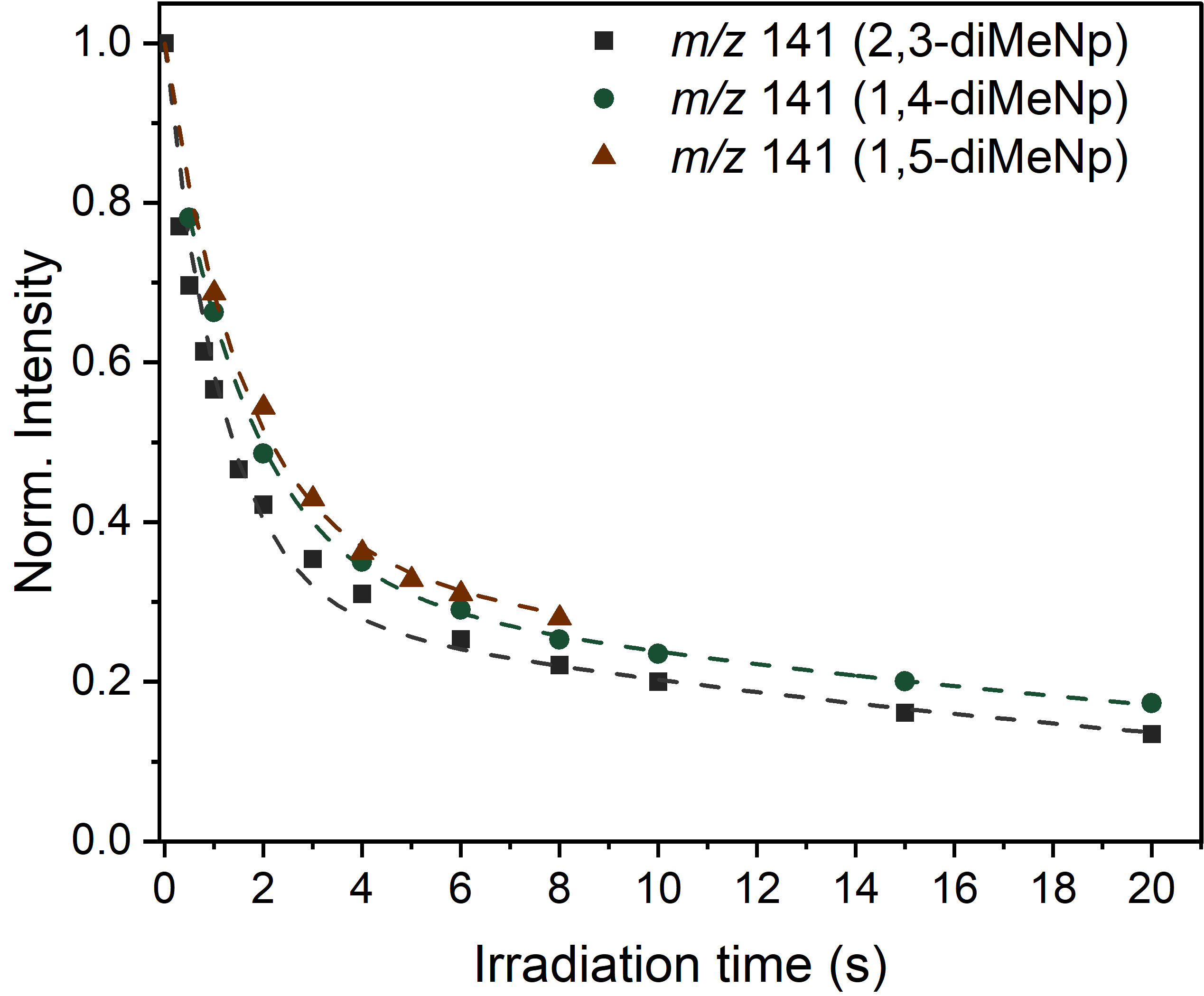}
    \caption{Photofragmentation kinetic curves of \ch{C11H9+} ions (produced from different precursors) under irradiation from the Xe arc lamp with a CF at 400 nm. Given the stability and quality of these measurements, we anticipate that the error bars will be minimal, generally within the size of the data markers. The experimental points were fitted using a biexponential decay model (dashed lines). The derived fitting parameters and the comparison with a  single exponential decay model are given in Table S1 and Figure S5, respectively.} 
    \label{CF400nm_141diMeNp}
\end{figure}

Figure~\ref{CF400nm_141diMeNp} shows the photofragmentation kinetics under irradiation with the Xe lamp of the \ch{C11H9+} species produced from all diMeNp investigated.
All curves, regardless of the precursor, display an initial fast decay followed by a slower one, quantified by a biexponential decay.

Such behavior was observed in our previous work on \ch{C11H9+} ions produced from 1-MeNp\cite{lozano2025}.
The distinct decay components were attributed to the coexistence of several long-lived isomers exhibiting distinct absorption features in the visible range  \cite{Nagy2011,lozano2025}. One could expect the presence of the same isomers for the \ch{C11H9+} species produced from diMeNp, given the structural similarities with MeNp.
This is further corroborated by the MD/DFTB simulations, as discussed in Section~\ref{MDsimulations}. The slower decay component could then be attributed to the dissociation of \ch{BzTr+}, as this isomer has the lowest absorption cross section in the visible range\cite{Nagy2011,lozano2025}.
Conversely, the faster decay likely arises from both \ch{NpCH2+} isomers, given their dominant photoabsorption in this spectral region\cite{Nagy2011,lozano2025}. However, the scenario may be further complicated by potential interconversion between isomers under photon irradiation.\cite{lozano2025}.

\subsubsection{Theoretical Insights}
\label{MDsimulations}

\begin{figure}[H]
    \centering
   \includegraphics[width=1\linewidth]{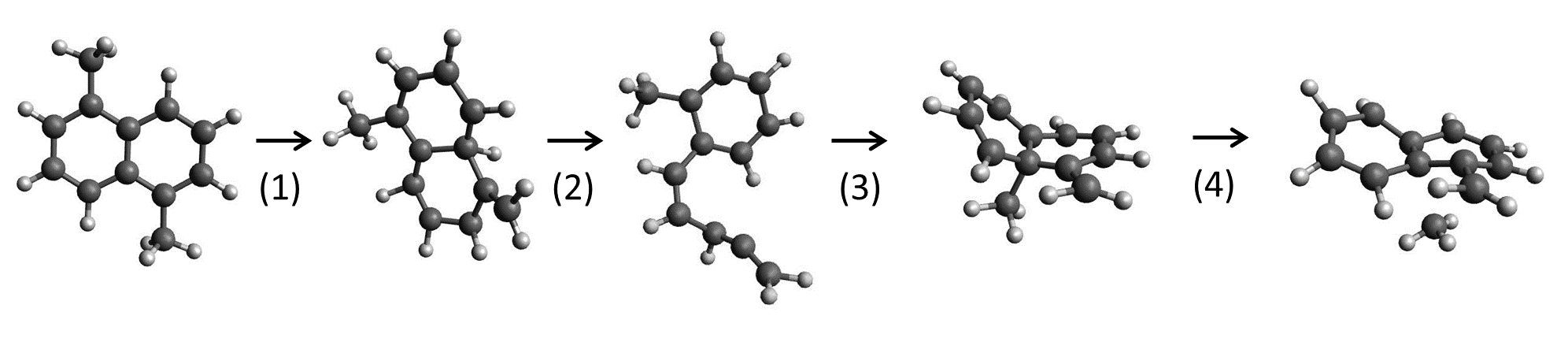}
    \caption{Snapshots from a 15.0~eV MD/DFTB simulation illustrating the dissociation of 1,5-\ch{diMeNp^{.}+} into 1-\ch{NpCH2+} + \ch{CH3}.}
    \label{snapshot_15diMeNp_1}
\end{figure}

To gain further insights into the dissociation dynamics of the three investigated diMeNp cations, we performed MD/DFTB simulations following the procedure described in Section~\ref{Theory}.
At the two examined energies (15.0 and 16.5~eV), \ch{CH3} loss is the dominant dissociation pathway, leading to the formation of different isomers of \ch{C11H9+}. The branching ratios (BRs) between the different isomers of \ch{C11H9+} are summarized in Tables~\ref{BR_MD_2ns} and S2.
The 1-\ch{NpCH2+} isomer is found to be the most abundant isomer when fragmenting 1,4- and 1,5-diMeNp cations at the two studied energies. The predominance of this isomer is even more striking in the case of 1,5-\ch{diMeNp^{.}+}. A representative trajectory of the MD simulations of 1,5-\ch{diMeNp+} resulting in 1-\ch{NpCH2+} is given in Figure~\ref{snapshot_15diMeNp_1}. It starts with a 1,3 H migration from the \ch{CH3} group to the closest fused carbon atom (step (1)). The introduction of an sp$^3$ carbon into the ring weakens the C–C bond, leading to ring opening (step (2)). This is followed by ring closure, facilitated by the electronic unsaturation of the free carbon chain (step (3)). In the intermediate formed, the \ch{CH3} group is attached to an sp$^3$ carbon within the ring, resulting in a weaker C–\ch{CH3} interaction compared to 1,5-\ch{diMeNp^{.}+}. Its dissociation restores aromaticity, yielding 1-\ch{NpCH2+}.

\begin{table}[!ht]
    \centering
    \begin{tabular}{|c||c|c|c|}
 \hline
     Precursor   & 1,4-diMeNp & 1,5-diMeNp & 2,3-diMeNp \\ 
\hline
    \#1-\ch{NpCH2+}/\#\ch{C11H9+}& 0.59 & 0.79 & 0.11 \\
\hline
    \#2-\ch{NpCH2+}/\#\ch{C11H9+}& 0.11 & 0.08 & 0.64  \\
\hline
    \#\ch{BzTr+}/\#\ch{C11H9+}& 0.17 & 0.04 & 0.15 \\
\hline
    \#Others/\#\ch{C11H9+}& 0.13 & 0.09 & 0.10 \\
\hline
    \end{tabular}
        \caption{Branching ratios for the dissociation of 1,4-, 1,5-, and 2,3-diMeNp cations, obtained from MD/DFTB simulations over a 2\,ns timescale at 15\,eV internal energy. The values were averaged over 180 simulations with randomly assigned velocity distributions. Note: \# denotes the number of occurrences.}
    \label{BR_MD_2ns}
\end{table}

\begin{figure}[H]
    \centering
   \includegraphics[width=1\linewidth]{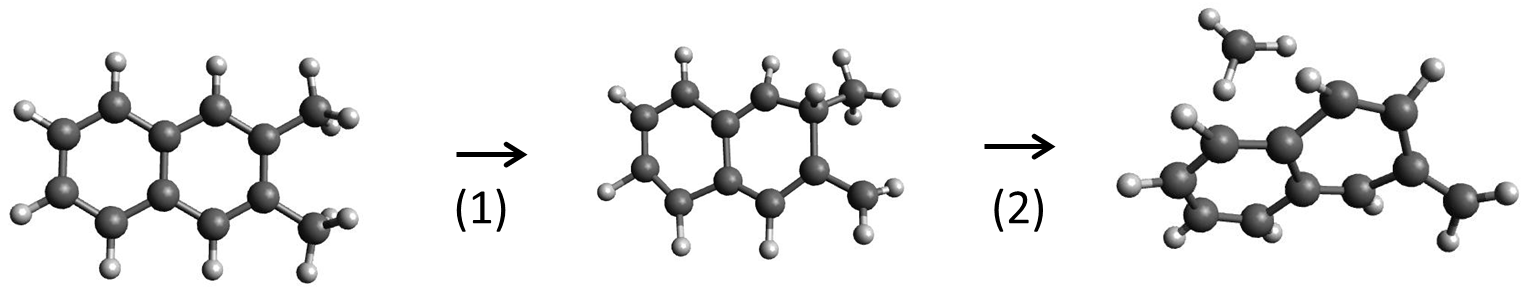}
    \caption{Snapshots from a 15.0~eV MD/DFTB simulation illustrating the dissociation of 2,3-\ch{diMeNp^{.}+} into 2-\ch{NpCH2+} + \ch{CH3}.}
    \label{snapshot_23diMeNp_2}
\end{figure}

\begin{figure}[H]
    \centering
   \includegraphics[width=1\linewidth]{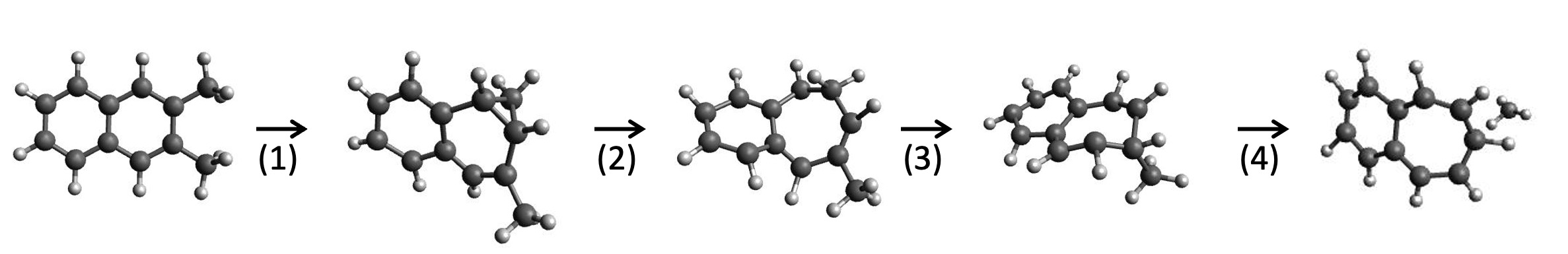}
    \caption{Snapshots from a 15.0~eV MD/DFTB simulation illustrating the dissociation of 2,3-\ch{diMeNp^{.}+} into \ch{BzTr+} + \ch{CH3}.}
    \label{snapshot_23diMeNp_BzTrop}
\end{figure}

As expected, the fragmentation of 2,3-\ch{diMeNp^{.}+} predominantly yields 2-\ch{NpCH2+} fragments, although a minor fraction of 1-\ch{NpCH2+} is also produced -- a situation that is comparable to the case of 1,4-\ch{diMeNp^{.}+}, but with the relative abundances of the \ch{NpCH2+} isomers inverted. A representative mechanism is shown in Figure~\ref{snapshot_23diMeNp_2}. As illustrated, the loss of \ch{CH3} is preceded by H migration, which generates an sp$^3$ carbon atom bonded to the \ch{CH3} group (step (1)). This weakens the C–\ch{CH3} bond, facilitating the subsequent loss of \ch{CH3}.
In addition, a significant fraction of \ch{BzTr+} is produced from 2,3- and 1,4-\ch{diMeNp^{.}+}, which is not the case for 1,5-\ch{diMeNp^{.}+}.
This can be rationalized by the expansion into a 7-membered ring as shown in Figure~\ref{snapshot_23diMeNp_BzTrop} for 2,3-\ch{diMeNp^{.}+}.

\subsection {Reactions between \ch{C11H9+} and dimethylated naphthalenes}
\label{reactivityresults}

In previous work\cite{lozano2025}, we observed the formation of \ch{C12H11+} at {\it m/z} 155 and an adduct \ch{C22H19+} at {\it m/z} 283 when \ch{C11H9+}, produced from 1-MeNp, reacted with its neutral precursor.
Here, we extend this investigation to the reactivity of \ch{C11H9+} species generated from diMeNp with their neutral precursors. In this case, the reaction products do not include any adduct but in addition to \ch{C12H11+} (eq~\ref{react155diMeNp}), another product, \ch{C13H13+} at {\it m/z} 169 (eq~\ref{react169diMeNp}) is observed. 
This product is particularly abundant in reactions involving 2,3-diMeNp (see Figure S2). The reaction channels observed at 300~K can therefore be summarized as:

\begin{align}
\begin{split}\label{react155diMeNp}
\ch[name-format=\sffamily\small]{!(\mz{141}) (C11H9^+) + C12H12 {}& -> !(\mz{155}) (C12H11^+) + C11H10}
\end{split}\\
\begin{split}\label{react169diMeNp}
\ch[name-format=\sffamily\small]{ {}& -> !(\mz{169}) (C13H13^+) + C10H8,}
\end{split}
\end{align}

The neutral products, although not detected,
likely correspond to methyl-naphthalene and naphthalene in reactions~\ref{react155diMeNp} and \ref{react169diMeNp}, respectively.

Table~\ref{tablereaction} summarizes the relative populations of reactive and nonreactive channels under complete reaction conditions (see Section~\ref{Experimental}). 
The BR between the two observed products varies notably depending on the specific diMeNp isomer. 
For 1,4- and 1,5-diMeNp, \ch{C12H11+} is the dominant product, whereas \ch{C13H13+} predominates in the case of 2,3-diMeNp.
These experimental findings can result from a combination of factors, including differences in the relative populations of the 1-\ch{NpCH2+} and 2-\ch{NpCH2+} isomers, as well as the reactivity of the neutral reactants, which may favor one reaction pathway over another.
In all cases, the non-reactive fraction, with relative abundances of 11–18\%, is attributed to \ch{BzTr+} based on our previous study\cite{lozano2025}. This trend is generally consistent with our MD simulations (Tables~\ref{BR_MD_2ns} and S2), although larger discrepancies are observed for 1,5-diMeNp compared to the other two precursors.

 \begin{table}[!ht]
    \centering
     \begin{tabular}{|c||c|c|c|}
    \hline
        Precursor/ & \multicolumn{2}{c|}{Reaction} & No Reaction \\
    \cline{2-4}
        Neutral reactant & \ch{C12H11+} & \ch{C13H13+} & \ch{C11H9+} \\
    \hline
        2,3-diMeNp & 0.07 & 0.79 & 0.14  \\
    \hline
        1,4-diMeNp & 0.70 & 0.12 & 0.18  \\
    \hline
        1,5-diMeNp & 0.83 & 0.06 & 0.11  \\
    \hline
     \end{tabular}
         \caption{Relative populations of both the reactive and nonreactive channels for reactions between \ch{C11H9+} and their neutral precursors.}
     \label{tablereaction}
 \end{table} 

\subsection {Photodissociation of reaction cationic products}\label{photoproducts}

We first focus on \ch{C12H11+} species, which are produced when \ch{C11H9+} ions react with the various diMeNp isomers, except for 2,3-diMeNp, where the relative abundance of \ch{C12H11+} was too low. This set of experiments also includes the \ch{C12H11+} product formed from 1-MeNp, prepared following the previously described procedure\cite{lozano2025}.

The photodissociation of all \ch{C12H11+} species under visible irradiation from the Xe lamp follows the same pathway, as illustrated by the representative mass spectrum in Figure S3. This involves the loss of either 2\ch{H} or \ch{H2}, with the resulting \ch{C12H9+} cation further dissociating through hydrogen loss to yield \ch{C12H8^{.}+} at {\it m/z} 152 (eq~\ref{dissmz155}).

\begin{align}
\ch[name-format=\sffamily\small]{!(\mz{155}) (C12H11^+)  -> [ - 2 H / H2] [h$\nu$] !(\mz{153}) (C12H9^+) -> [ - H] [h$\nu$] !(\mz{152}) (C12H8^+)}
\label{dissmz155}
\end{align}

\begin{figure}[H]
    \centering
   \includegraphics[width=1\linewidth]{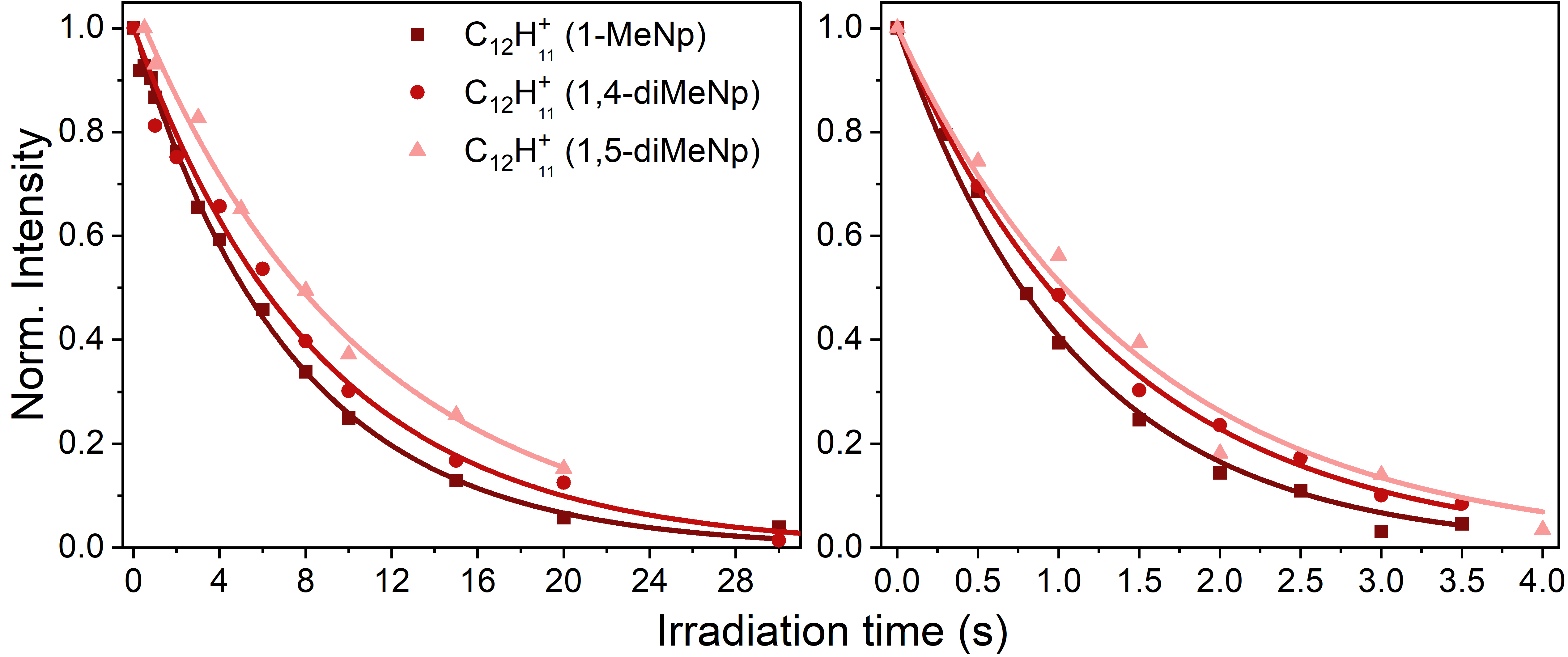}
    \caption{Photofragmentation kinetic curves of \ch{C12H11+} ions under irradiation from the Xe arc lamp with two different filter configurations. Left panel: CF at 400~nm + NDF. Right panel: CF at 400~nm alone. Solid lines represent exponential decay fits (Table S1). The \ch{C12H11+} ions were generated from different precursors (see text for further details).}
    \label{2conditions_155}
\end{figure}

\begin{figure}[!ht]
    \centering
   \includegraphics[width=1.0\linewidth]{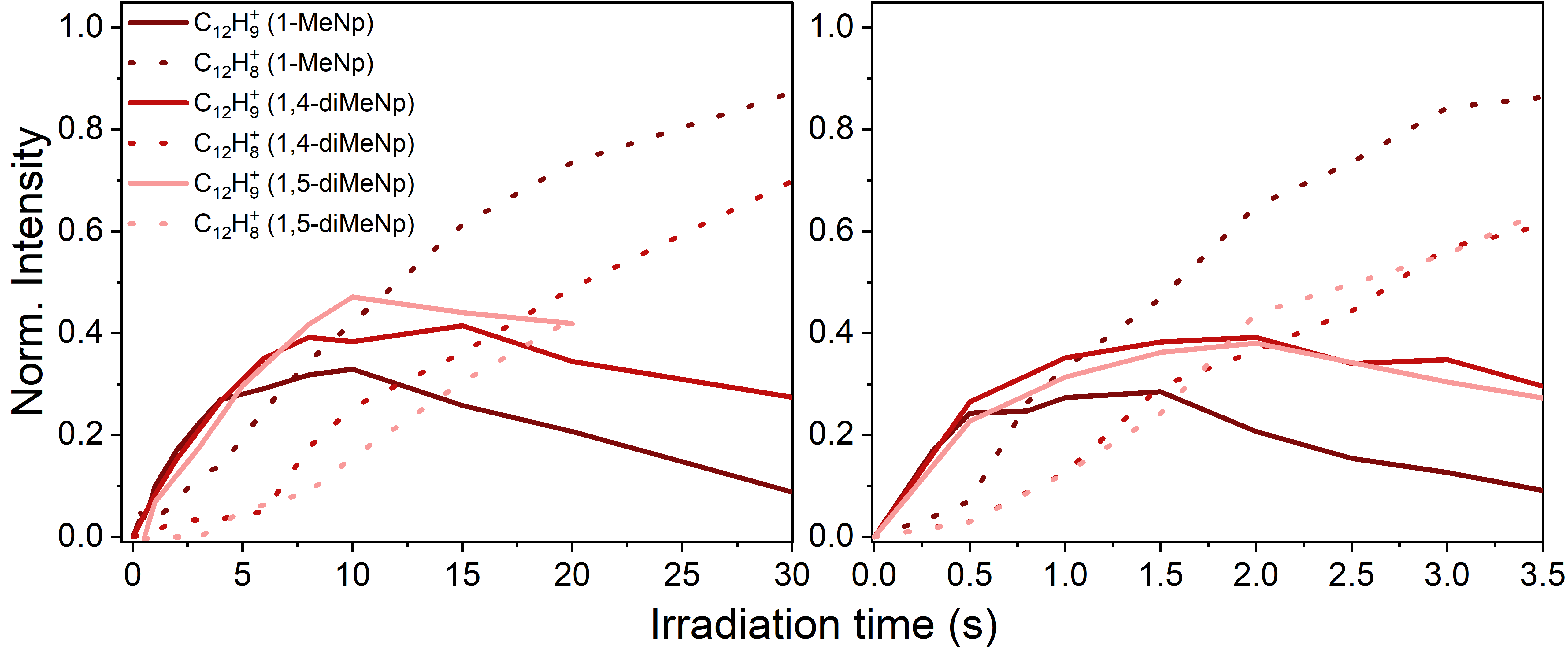}
    \caption{Normalized fragment intensities from the photodissociation kinetics of \ch{C12H11+} under irradiation with the Xe arc lamp. Left panel: CF at 400~nm + NDF. Right panel: CF at 400~nm alone.}
    \label{mz153mz152}
\end{figure}

Figure~\ref{2conditions_155} shows the photofragmentation kinetics of \ch{C12H11+} species recorded under broadband irradiation for $\lambda \gtrsim 400$~nm. Slight differences are observed in the decay of the parent ions. The \ch{C12H11+} species formed from 1-MeNp exhibit the fastest decay, whereas those formed from 1,5-diMeNp show, under the lower irradiation conditions, a transient delay (or induction time\cite{UechiDunbar1992}) of approximately 1.5~s before the exponential decay.
This behavior is expected in a regime where photon absorption heating competes with radiative cooling\cite{UechiDunbar1992,Boissel1997}. Specifically, under Xe lamp irradiation, an internal energy distribution is established, and dissociation is primarily driven by single-photon absorption in ions with the highest internal energy.\cite{Boissel1997}. Thus, the transient delay in dissociation corresponds to the time required to establish this internal energy distribution, which occurs under conditions of low heating efficiency and, consequently, low visible-range absorption cross sections. This points to different isomers being involved with different absorption cross sections. A similar reasoning applies to the dissociation of the primary fragment \ch{C12H9+} at {\it m/z} 153 (Figure~\ref{mz153mz152}).

\begin{figure}[!ht]
    \centering
   \includegraphics[width=1.0\linewidth]{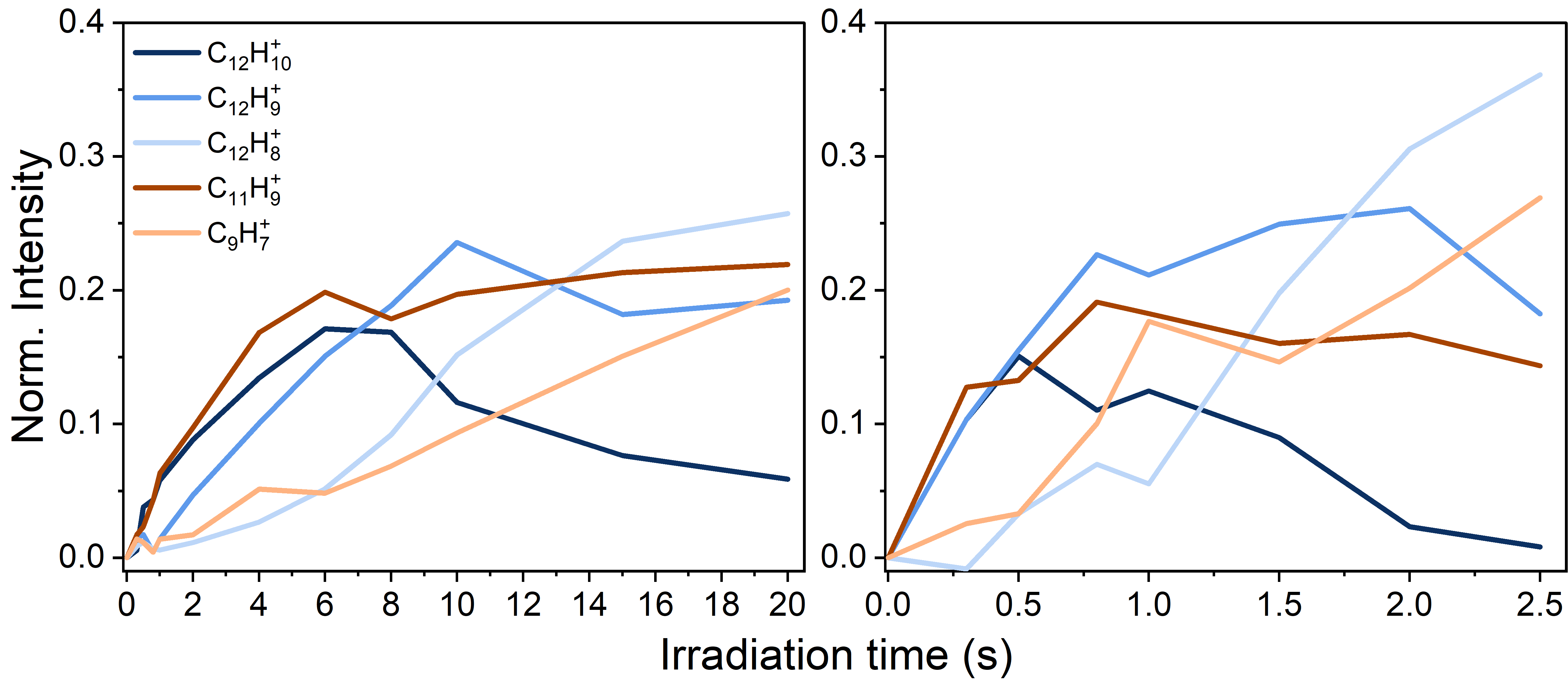}
    \caption{Normalized fragment intensities from the photodissociation kinetics of \ch{C13H13+} species under two irradiation conditions. Left panel: CF at 400~nm + NDF. Right panel: CF at 400~nm alone.}
    \label{mz169fragm}
\end{figure}

The \ch{C13H13+} products were studied for the reaction of \ch{C11H9+} with 2,3-diMeNp. A representative mass spectrum after irradiation is shown in Figure S3, and the dissociation kinetics are presented in Figure S4. Detailed kinetics curves (Figure~\ref{mz169fragm}) reveal two primary fragments, \ch{C12H10^{.}+} and \ch{C11H9+}, with branching ratios of around 60\% and 40\%, respectively, independent of photon flux.
The first dissociation channel involves a mass loss of 15~u, consistent with \ch{CH3} elimination (Eq~\ref{mz169mz154}). The resulting \ch{C12H10^{.}+} then undergoes successive H losses (Figure~\ref{mz169fragm}) to form \ch{C12H8^{.}+}. The second channel proceeds via a 28~u (\ch{C2H4}) loss, yielding \ch{C11H9+} cations, which can further dissociate into \ch{C9H7+} and neutral \ch{C2H2} (Eq~\ref{mz169mz141}).
The photofragmentation channels are summarized as follows:

\begin{align}
\begin{split}\label{mz169mz154}
\ch[name-format=\sffamily\small]{!(\mz{169}) (C13H13^+) {}& -> [ - CH3] [h$\nu$] !(\mz{154}) (C12H10^{.+}) -> [ - H] [h$\nu$] !(\mz{153}) (C12H9^+) -> [ - H] [h$\nu$] !(\mz{152}) (C12H8^{.+}) }
\end{split}\\
\begin{split}\label{mz169mz141}
\ch[name-format=\sffamily\small]{{}& -> [ - C2H4][h$\nu$] !(\mz{141}) (C11H9^+) -> [ - C2H2][h$\nu$] !(\mz{115}) (C9H7^+) }
\end{split}
\end{align}

We further investigated the \ch{C13H13+} species using MPD spectroscopy (see Section~\ref{Experimental}). 
The recorded MPD spectrum is shown in Figure S7, displaying both the total dissociation yield and the contributions from the individual fragmentation channels. 
Overall, dissociation is observed across the entire 420-600\,nm range. The broad spectral profile suggests that isomerization effects may be involved, as previously observed in our earlier study\cite{lozano2025}. The branching ratio for the two dissociation channels of \ch{C13H13+} was determined (Figure S8). Its average values is consistent with the values of 60 and 40\% obtained for the \ch{C12H10^{.}+} and \ch{C11H9+} channels, respectively, under broad-band excitation with the Xe lamp (Figure S3). The analysis of the secondary fragments (Figure S7) reveals trends despite a relatively low (and noisy) signal:
(i) \ch{C12H10^{.}+} shows negligible dissociation in the 450–550 nm range and absorbs at shorter and longer wavelengths, as shown by the signal in \ch{C12H8^{.}+} and \ch{C12H9+}, respectively ;
(ii) There is evidence that \ch{C11H9+} dissociates in the 510--580 nm range, producing \ch{C9H7+}. This is supported by the averaged value of the \ch{C11H9+} signal, which is 0.01$\pm$0.01 in the 420-500~nm range compared to 0.058$\pm$0.016 in the 510-580~nm range.

\begin{figure}[H]
    \centering
    \includegraphics[width=0.7\linewidth]{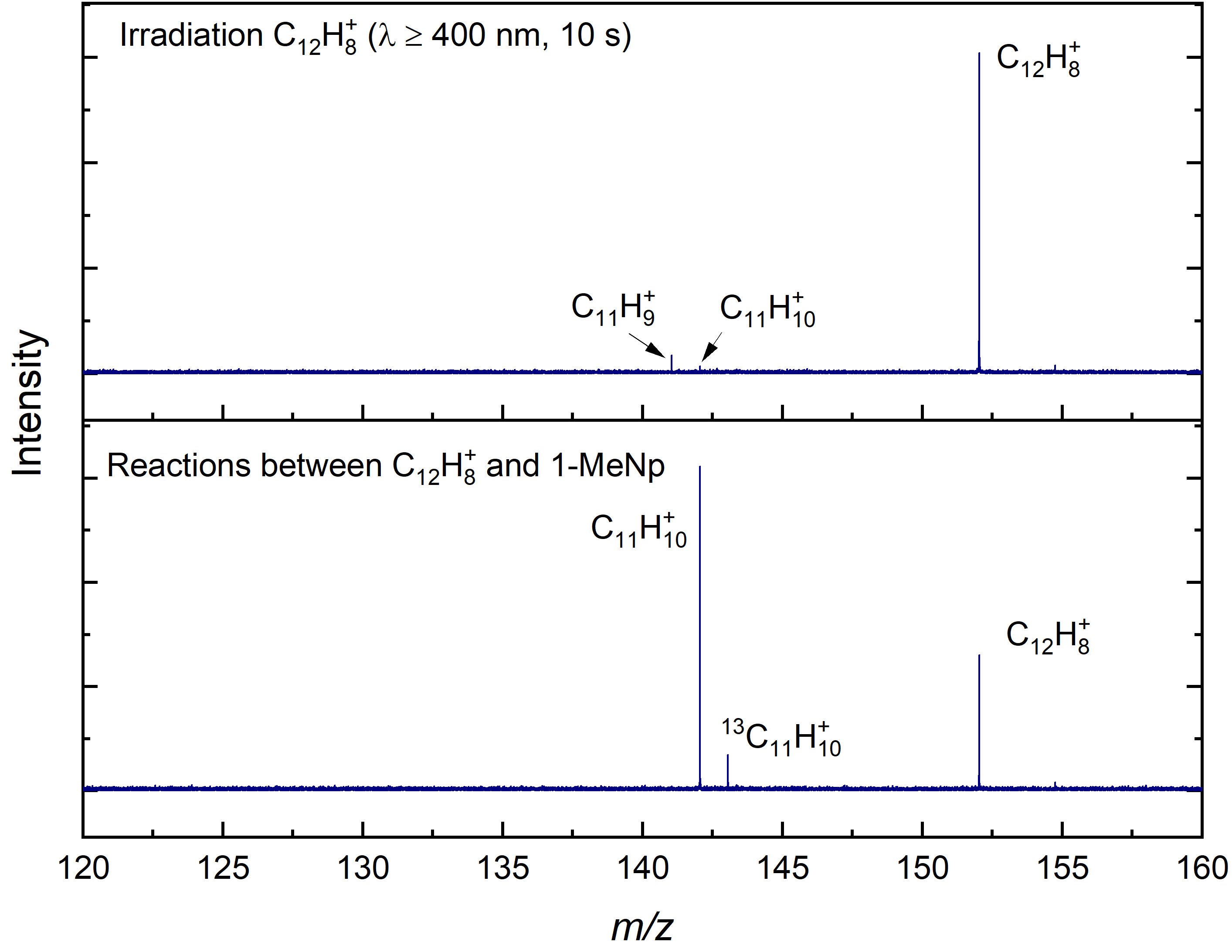}
    \caption{Top panel: mass spectrum recorded after broadband irradiation (10~s, $\lambda$ $\gtrsim$ 400~nm) of the \ch{C12H8^{.}+} species. Bottom panel: mass spectrum recorded after reactions (charge transfer) between \ch{C12H8^{.}+} and neutral 1-MeNp. In both cases, the \ch{C12H8^{.}+} species were generated from the 1-MeNp precursor.
    The minor peaks in the upper panel arise from charge transfer involving residual neutral precursor in the cell, producing \ch{C11H10^{.}+}, which further dissociates into \ch{C11H9+} upon Xe lamp irradiation.}
    \label{mz152MS}
\end{figure}

Finally, because \ch{C12H8^{.}+} species was identified as a common final photofragment of both \ch{C12H11+} and \ch{C13H13+}, we examined this species in more detail.
To this end, we isolated the \ch{C12H8^{.}+} cations produced via Eqs~\ref{dissmz155} and \ref{mz169mz154}, using parent cations generated from 1-MeNp, 1,4-diMeNp, and 2,3-diMeNp.
In all cases, \ch{C12H8^{.}+} was found to be remarkably photostable compared with the other studied species, both under visible broadband irradiation and under narrowband irradiation at 266~nm.
Moreover, \ch{C12H8^{.}+} was observed to undergo charge-transfer reactions with 1-MeNp and with  all investigated diMeNp isomers. Figure~\ref{mz152MS} presents two representative spectra illustrating both the photostability and the charge-transfer reactions of \ch{C12H8^{.}+} produced from parent cations generated from 1-MeNp. 

\section{Discussion}
\label{Discussion}
The results presented in Section~\ref{results} establish \ch{C11H9+} as a key intermediate in the photodissociation and subsequent reactivity of methylated naphthalenes.
Its prominence as a common photofragment of both diMeNp and MeNp, together with its ability to react with these species, leads to the formation of two primary reaction products, \ch{C12H11+} and \ch{C13H13+}. Photophysical characterization of these two products further reveals distinct fragmentation pathways.

In the following sections, we examine how our experimental findings data support the formation of acenaphthene derivatives and can be used to propose chemical scenarios that could account for the production of these species.

\subsection{Evidence for acenaphthene derivatives}
\label{Discuss_acenaph}
\begin{figure}[ht]
    \centering
   \includegraphics[width=0.8\linewidth]{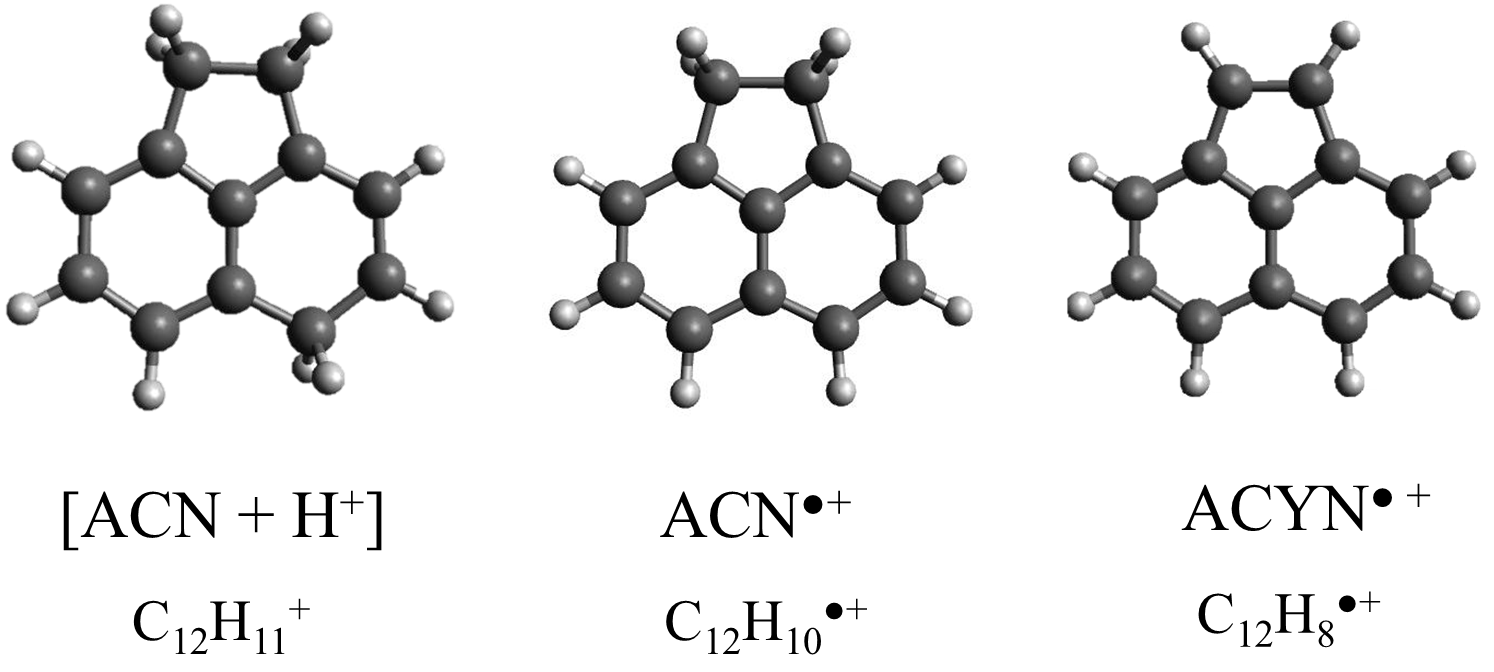}
    \caption{Molecular structures of the most stable isomers of protonated acenaphthene ([ACN+\ch{H+}]), acenaphthene radical cation (\ch{ACN^{.}+}), and acenaphthylene radical cation (\ch{ACYN^{.}+}).}
    \label{molecules_acenaphthenes}
\end{figure}

Among the candidate structures for the \ch{C12H11+} products, our results point to protonated acenaphthene, [ACN+\ch{H+}] (Figure~\ref{molecules_acenaphthenes}).
This assignment is supported by dissociation via 2H/\ch{H2} loss, consistent with the results reported by \citet{szczepanski2010}, who showed that [ACN+\ch{H+}] dissociates primarily through the ejection of two mass units, attributed to \ch{H2} loss from the pentagonal ring.

Regarding the \ch{C13H13+} products, our experimental results provide evidence for assigning its two primary fragments, \ch{C12H10^{.}+} and \ch{C11H9+}.
The trends observed in the MPD spectrum are consistent with \ch{C12H10^{.}+} being the acenaphthene radical cation, \ch{ACN^{.}+} (see Figure S7 and Section~\ref{photoproducts}). This proposal is supported by previous studies of the acenaphthene radical cation (\ch{ACN^{.}+})\cite{Boschi1974, Andrews1985, shida1988, Banisaukas2003} and by our computed vertical excitation energies (see Table S3). Moreover, the observed sequential loss of hydrogen atoms aligns with earlier reports on the photodissociation of \ch{ACN^{·}+}\cite{Banisaukas2003}. In addition, the observation of \ch{C9H7+} in the 510-580\,nm range reveals that this ion is produced by the dissociation of 2-\ch{NpCH2+}, which is the long-lived isomer of \ch{C11H9+} absorbing in this spectral range.\cite{Nagy2011,lozano2025}

Additional experimental results support the formation of acenaphthene derivatives. We found that the \ch{C12H8^{.}+} species, produced as the common final fragment from the dissociation of both \ch{C12H11+} (eq~\ref{dissmz155}) and \ch{C12H10^{.}+} (eq~\ref{mz169mz154}), undergo charge-transfer reactions with 1-MeNp as well as with 1,4- and 2,3-diMeNp.
This behavior is consistent with \ch{C12H8^{.}+} corresponding to the acenaphthylene radical cation (\ch{ACYN^{.}+}), as the ionization potential of ACYN is higher than that of all compounds with which charge-transfer was observed\cite{NISTChemistryWebBook}.
Furthermore, \ch{ACYN^{.}+} is known to be photostable under broadband Xe-lamp irradiation\cite{Ekern1998}, in agreement with our observations.

In summary, reactions between \ch{C11H9+} and methylated naphthalenes yield two species associated with acenaphthene derivatives: one likely formed directly (\ch{C12H11+}) and one produced through subsequent photoprocessing of the \ch{C13H13+} products (\ch{C12H10^{.}+}).
Upon irradiation, both species ultimately converge to the stable \ch{C12H8^{.}+}, which is most consistent with the acenaphthylene cation.

\subsection{Formation of acenaphthene derivatives: building chemical scenarios}
\label{Discuss_scenarios}
We outline in this section plausible key steps involved in the reactions between \ch{C11H9+} and methylated naphthalenes. While the proposed chemical scenarios are supported by our DFT calculations, reaching a definitive conclusion would require more extensive computational investigation of the reaction pathways. Further details on the calculations performed in this study are provided in Section S7.

We first consider the formation of [ACN+\ch{H+}] with 1-\ch{NpCH2+} + 1-MeNp as reactants. The underlying hypothesis is that [ACN+\ch{H+}] formation involves C-C coupling between the methylene carbon of the molecular cation and the methyl carbon of the neutral precursor. This coupling requires electronic unsaturation in the -\ch{CH3} group, which can be achieved by H migration, either within the same molecule or between molecular units, from the methyl group to the closest carbon atom of the PAH skeleton.

\begin{figure}[H]
    \centering
   \includegraphics[width=1\linewidth]{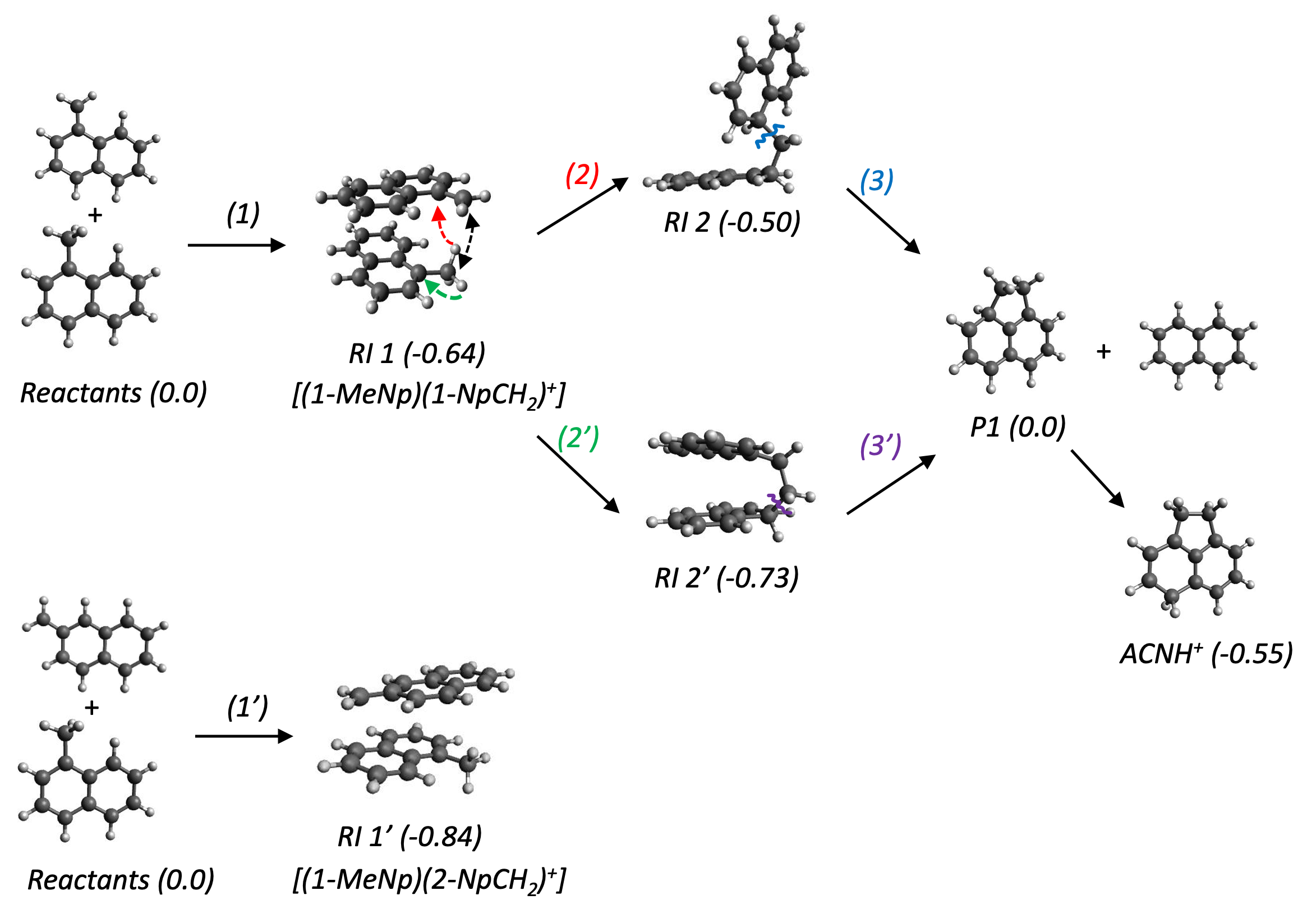}
    \caption{Geometries of the reactants, products and possible reactive intermediates (RI) involved in the reactions of 1-MeNp with 1-\ch{NpCH2+} (top) and 2-\ch{NpCH2+} (bottom), as determined at the B3LYP-GD3\cite{Becke1993,Grimme2010}/6-31G(d,p) level of theory (see Section S7 for computational details).  Relative enthalpies at 298\,K, with respect to the reactants, determined by B3LYP-GD3/aug-cc-pvtz//B3LYP-GD3/6-31G(d,p) single-point calculations, are indicated in parenthesis (in eV).}
    \label{fig:schemeC12H11+}
\end{figure}

The proposed mechanisms are sketched through three main steps as depicted at top of  Figure~\ref{fig:schemeC12H11+}: (1) formation of an ion-molecule complex RI~1 located 0.64 eV below the reactants at 298\,K. (2) or (2') H migration in which the H of the -\ch{CH3} group migrates to the closest C atom of the other partner (intermolecular shift leading to step (2)) or to the closest C atom on the same molecule (intramolecular shift ; more specifically 1,2H migration leading to step (2')). These hydrogen transfers are accompanied by C-C coupling between the carbons of the two methylene groups, and this results in the formation of RI~2 or RI~2', which are located 0.50 and 0.73\,eV  below the reactants, respectively. This step ((2) or (2')) is favored by the proximity of the two molecules in the ion-molecule complex. (3) or (3') Breaking of the  -HC(sp$^3$)-H$_2$C(sp$^3$)- bond (originally -C=CH$_2$$^+$ bond in step (2) and -C-CH$_3$ bond in step (2'), weakened after hydrogenation of the originally sp$^2$ carbon atom and C-bonding of the other carbon atom) leading to the formation of \ch{C12H11+}, the precursor of one isomer of [ACN+\ch{H+}] (P1 on Figure~\ref{fig:schemeC12H11+}) through formation of a 5-carbon ring by closure of the \ch{-CH2–CH2^.} chain, and loss of neutral naphthalene. Energetically, P1 +Nap is iso-energetic to the reactants. Subsequent H shift would lead to the most stable isomer of [ACN+\ch{H+}], found -0.55\,eV below the reactants.

In the case of  2-\ch{NpCH2+} + 1-MeNp as reactants, the geometry of the initial (2-\ch{NpCH2+})(1-MeNp) complex is expected to be different (see RI~1' at the bottom of Figure~\ref{fig:schemeC12H11+}). From the optimization procedure described in Section S7, we identified a stable complex (-0.84\,eV) below the reactants, in which the methylene and methyl carbon atoms do not face each other. This configuration is unfavorable for reactivity, likely explaining the presence of the \ch{C22H19+} adduct observed in our earlier study\cite{lozano2025}.

All our results therefore point to ring closure as the mechanism leading to the formation of \ch{C12H11+} ({\it m/z} 155). Our experiments show no evidence of \ch{CH3} loss or any other C-loss channel in the dissociation of {\it m/z} 155. In addition, MD simulations of the two \ch{C12H11+} isomers featuring a \ch{=CH-CH3} chain at positions 1 and 2 (see Section S6) show no significant formation of [ACN+\ch{H+}], in contrast with the experimental identification of \ch{C12H11+} as [ACN+\ch{H+}]. If ring closure is not favored for 2-\ch{NpCH2+}, the \ch{C12H11+} product at {\it m/z} 155 would result specifically from the reactivity of 1-\ch{NpCH2+}with 1-MeNp, yielding [ACN+\ch{H+}] without additional energy input to the formed complex.

In this context, the differences observed in the photodissociation kinetics of both \ch{C12H11+} and \ch{C12H9+} species, depending on the neutral precursor (see Figures~\ref{2conditions_155} and \ref{mz153mz152}), can be rationalized by variations in the position of the additional H atom on the hexagonal rings.
For the \ch{C12H9+} species, we calculated the bond dissociation energies of the extra H at positions 1 to 5 (Section S5). Comparable energies are found for the three positions on the hexagonal rings, indicating that the differences observed in the kinetic curves (see Figure~\ref{mz153mz152}) most likely arise from variations in the visible absorption cross sections.

\begin{figure}[H]
    \centering
   \includegraphics[width=0.5\linewidth]{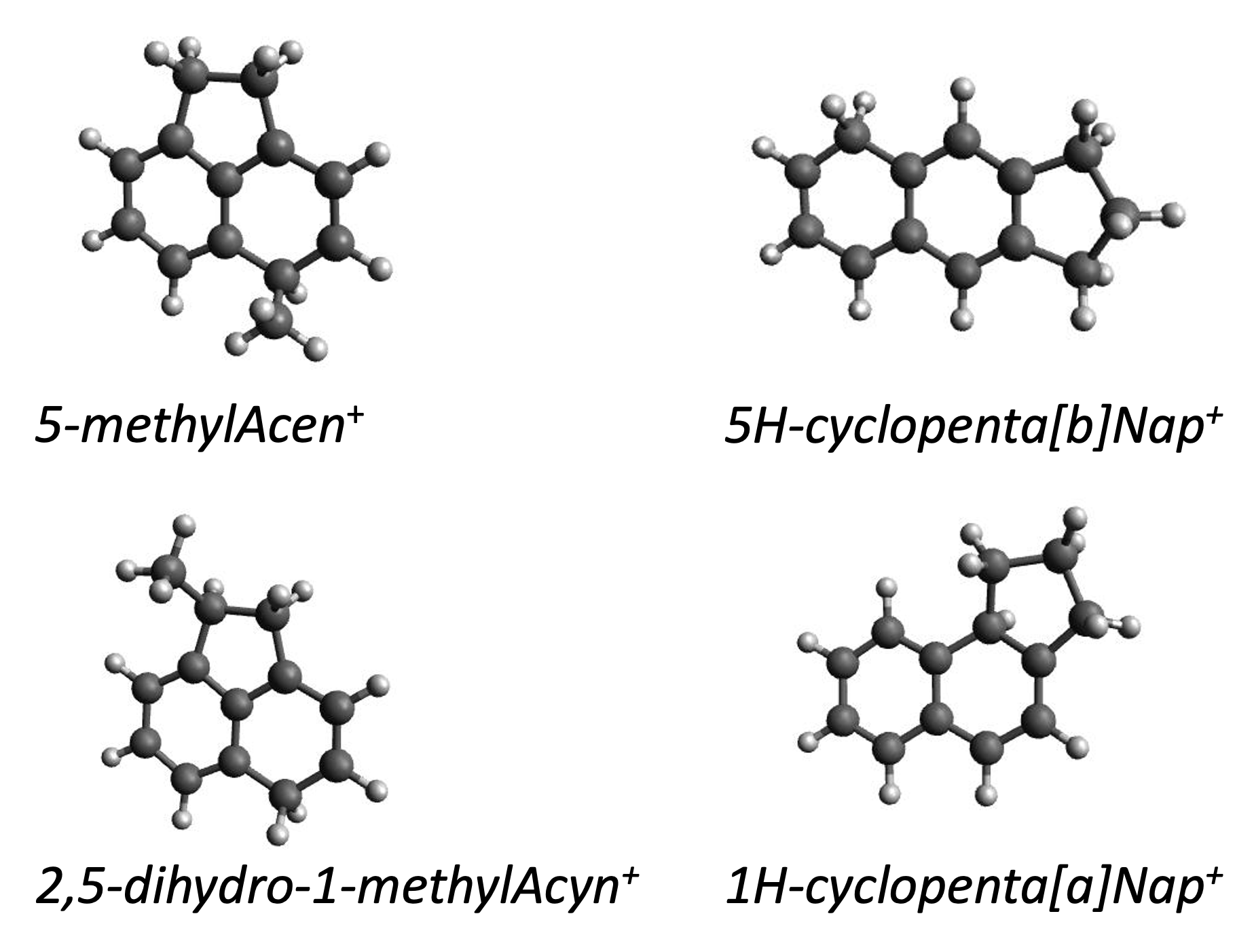}
    \caption{Structures of \ch{C13H13+} isomers likely formed in the experiments, selected from several energetically accessible possibilities (see Figure S10). These structures were determined at the B3LYP-GD3/6-31G(d,p) level of theory. Acen : acenaphthene, Nap : naphthalene, Acyn : acenaphtylene.}
    \label{fig:schemeC13H13+}
\end{figure}

The reactivity products of \ch{C11H9+} with diMeNp can also be rationalized by following the steps outlined in Figure~\ref{fig:schemeC12H11+}. The presence of two methyl groups in the neutral reactant can lead to different isomers, depending on the specific diMeNp isomer involved. The 2,3-diMeNp would favor the formation of 5H-cyclopenta[b]Nap$^+$ (Figure~\ref{fig:schemeC13H13+}) following initial intermolecular H shift - and subsequent intramolecular H shift to allow ring closure. This \ch{C13H13+} isomer lies 0.34\,eV below the reactants (see Table S6 and Figure S10). 
Starting from the 1,5- and 1,4-diMeNp neutral precursors, the formation of  \ch{C13H13+} is observed to be a minor channel in our experiments. In this case, the produced isomer could be 5-methylAcen$^+$ (see Figure~\ref{fig:schemeC13H13+}), formed through an initial intermolecular H-shift mechanism as in Figure~\ref{fig:schemeC12H11+} (steps (2) and (3)), considering that H-shifts occur afterwards to lead to the most stable isomer. However, the major reaction pathway is found to be \ch{C12H11+}, which is consistent with a mechanism initiated by an intramolecular H shift in diMeNp, similar to that sketched in Figure~\ref{fig:schemeC12H11+} (steps (2') and (3')).

Finally, we must consider the photoproducts of \ch{C13H13+} generated from 2,3-diMeNp. Upon heating with the Xe lamp, the 5H-cyclopenta[b]Nap$^+$ structure could undergo chain opening, promoting \ch{C2H4} loss and yielding 2-\ch{NpCH2+}. This is consistent with experimental results (eq~\ref{mz169mz141} and point (ii) in the analysis of the MPD spectrum in Section~\ref{photoproducts}). Alternatively, chain opening could lead to a new isomer after ring closure, forming 1H-cyclopenta[a]Nap$^+$ (Figure~\ref{fig:schemeC13H13+}). A subsequent chain opening/closing step could produce 2,5-dihydro-1-methylAcyn$^+$ (Figure~\ref{fig:schemeC13H13+}), a candidate for \ch{CH3} loss and the formation of \ch{C12H10^{.}+}, an acenaphthene derivative. In our experiments, the latter represents a major dissociation channel, accounting for 60\% of the \ch{C13H13+} dissociation products.

\section{Implications}

The chemistry discussed in this work is centered on the reactivity of the \ch{C11H9+} ion, a common fragment produced during the dissociative ionization (or sequential ionization and dissociation) of methyl- and dimethyl-naphthalene. Species in which one hydrogen atom of a PAH is replaced by a methylene cation (\ch{CH2+}) form a family of carbocations that are expected to be highly reactive\cite{Olah1995}. In addition, they are resonance-stabilized through charge delocalization over the aromatic $\pi$-system. Their study is of interest not only for fundamental gas-phase chemistry, but also for understanding processes in astrochemical environments such as cold interstellar clouds and Titan’s ionosphere.

Early studies on the benzylium/tropylium (\ch{C7H7+}) system showed that benzylium reacts with toluene to form \ch{C8H9+}\cite{Shen1974,Ausloos1980}, via a proposed mechanism involving \ch{CH2+} transfer from the ion to the neutral molecule without H scrambling (see Scheme~II in Shen \& Dunbar\cite{Shen1974}). 
This mechanism alone is insufficient to explain all our experimental results. Instead, we propose an alternative mechanism based on the formation of a long-lived complex (adduct), which can facilitate conformer exploration, thereby promoting H migration, C–C coupling, and, in most cases, the formation of a pentagonal cycle.
Our calculations also suggest that the geometry of the (2-\ch{NpCH2+})(1-MeNp) complex is less favorable for C-C coupling. A more detailed exploration of the ion-molecule potential energy surface (PES) would allow to conclude on the attribution of the \ch{C22H19+} species to this complex.

Two isomers of cyanoacenaphthylene (1- and 5-\ch{C12H7CN}) were recently detected in TMC-1\cite{Cernicharo2024}. Interestingly, the derived column density of this species is 1.7 times higher than that of the smaller cyanonaphthalene isomers (1- and 2-\ch{C10H7CN})\cite{McGuire2021}. Phenalene (\ch{C13H10}) has also been detected in TMC-1\cite{Cabezas2025}. The authors of that study proposed that phenalene is most likely formed through the reaction of acenaphthylene (ACYN) with \ch{CH3+}, producing \ch{C13H11+} via radiative association, followed by dissociative recombination to yield \ch{C13H10}. However, no pathway was proposed for the formation of ACYN itself, highlighting missing reactions in current chemical networks that could enable efficient PAH growth in cold dark clouds\cite{Cernicharo2024}.

The chemistry presented here reveals an efficient route to 
acenaphthene derivatives through ion–molecule reactions, which are expected to proceed efficiently at low temperatures. Although the experiments were conducted at 300~K rather than under cryogenic conditions, the isolation conditions ensure that only radiatively stabilized processes occur, mirroring those relevant in astrophysical environments. The key step involves the reactivity of \ch{C11H9+}, a typical fragment arising from dissociative ionization of methyl- and dimethyl-naphthalene and its complexation with these neutral species favoring CC coupling. Photoprocessing of the two reactivity products, \ch{C12H11+} and \ch{C13H13+} leads to the formation of the acenaphthylene radical cation, \ch{ACYN^{.}+}.
In astrophysical environments, additional energy input could be supplied by mild UV photons or electron recombination.

Jochims et al. reported a critical energy of 7.8 eV for the survival of \ch{ACYN^{.}+} in astrophysical environments, compared to only ~3.9 eV for methyl-naphthalene cations\cite{Jochims1999}. More recent experiments using electrostatic storage rings have enabled measurements of improved astrophysical relevance by probing the competition between radiative cooling and dissociation of isolated ions over long timescales\cite{Martin2015, Stockett2020}. Using the cryogenic DESIREE setup, it was recently determined that \ch{ACYN^{.}+} remains stable up to vibrational energies of approximately 7.6 eV\cite{Subramani25}, a value comparable to that of the anthracene cation (\ch{C14H10+})\cite{Martin2015} and significantly higher than the 5.8 eV derived for naphthalene cations\cite{Lee2023}. \ch{ACYN^{.}+} was identified as the main fragment ion produced in the dissociative ionization of \ch{C14H10} (anthracene and phenanthrene)\cite{Banhatti2022}, which indicates that it is also the principal fragment of \ch{C14H10+} dissociation. However, any environment capable of photodissociating \ch{C14H10+} would also lead to the dissociation of \ch{ACYN^{.}+}, making it unlikely that this pathway accounts for the cyanoacenaphthylene observed in TMC-1.

The chemistry explored in our study may also influence hydrocarbon chemistry in dense and FUV-irradiated regions, such as the Orion Bar. These photodissociation regions (PDRs) contain dense and warm gas (T$_{\rm gas}$ = 300–600~K) and exhibit active chemistry of small hydrocarbon radicals, such as \ch{C2H} and \ch{CH2}\cite{Goicoechea2025}. Models predict abundances of these small hydrocarbons at their peak of a few 10$^{-7}$ relative to H. Assuming that 10\% of carbon is locked in PAHs with a C/H ratio of 1.4 × 10$^{-4}$, the PAH abundance would be 1.4 × 10$^{-6}$ if all PAHs are similar in size to naphthalene, or 2.8 × 10$^{-7}$ for PAHs with an average size of 50 C atoms. Under these conditions, highly reactive radicals could readily react with PAHs, potentially enabling the type of chemistry explored in this study.

\ch{C11H9+} is among the abundant large cations observed by \textit{in situ} mass spectrometry in Titan’s ionosphere\cite{Crary2009, Haythornthwaite2021} and could therefore play a key role in PAH chemistry there\cite{Ali2015}. Understanding this chemistry is important, as it contributes to the formation of tholins, which are responsible for Titan’s atmospheric hazes\cite{Waite2007}. Westlake et al.\cite{Westlake2014} showed that Cassini mission data on large organic ions are consistent with ion–molecule reactions, highlighting the formation of \ch{C6H7+} as a precursor of benzene. They also emphasized the need to investigate ionic chemistry for larger species, particularly those containing 8–13 carbon atoms\cite{Westlake2014,Loison2019}. Our results on the reactivity of \ch{C11H9+} are directly relevant to Titan’s ionosphere, indicating an active photochemistry that promotes the growth of PAHs, including species containing pentagonal rings.

\section{Conclusion}

Using the multiplex capabilities of PIRENEA, we studied the reactivity of \ch{C11H9+} with methylated naphthalenes under isolated conditions relevant to gas-phase chemistry in space\cite{Herbst2021}. The \ch{C11H9+} ions at {\it m/z} 141 were consistently identified as key fragments produced in the photodissociation of MeNp and diMeNp cations. These ions comprise three long-lived \ch{C11H9+} isomers -- 1-\ch{NpCH2+}, 2-\ch{NpCH2+}, and \ch{BzTr+} -- previously identified from 1-MeNp and 2-MeNp\cite{lozano2025}.

The benzylium-like isomers of \ch{C11H9+} react with neutral MeNp and diMeNp to form two primary products at {\it m/z}~155 and {\it m/z}~169, with relative abundances that depend on the precursor. The corresponding reaction products, \ch{C12H11+} and \ch{C13H13+}, exhibit distinct photodissociation pathways: \ch{C12H11+} predominantly undergoes sequential hydrogen loss, whereas \ch{C13H13+} follows two competing channels, yielding either \ch{C12H10^{.}+} or regenerating \ch{C11H9+}. In both cases, \ch{C12H8^{.}+} emerges as a stable terminal photofragment upon irradiation with visible light from the Xe lamp.

We propose a chemical scenario involving long-lived ion-molecule complexes, where conformer exploration can facilitate C–C coupling and the formation of a pentagonal cycle. While the proposed reaction pathways require further investigation to fully validate each step, our experimental results, supported by DFT calculations, demonstrate that this chemistry offers a viable mechanism for the formation of acenaphthene derivatives in TMC-1, for which no formation pathway has yet been identified\cite{Cernicharo2024}.

Our study highlights the central role of \ch{C11H9+} as both a common photofragment and a reactive intermediate that drives the formation of acenaphthene derivatives and related species. This does not, however, rule out alternative growth pathways involving the reactivity of naphthalene with small hydrocarbon radicals and their ions, which would require dedicated investigations. We are entering an exciting era in which the carbon chemistry underlying chemical complexity in star- and planet-forming regions can be characterized in ever greater detail using powerful astronomical instruments\cite{Goicoechea2025}. The discovery of PAHs in cold clouds with radio telescopes, together with the recent detection of \ch{CH3+} by JWST\cite{Berne2023}, poses new challenges for chemical modeling and opens perspectives for laboratory astrophysics in the coming years.  These experiments must nevertheless contend with the unusual conditions prevailing in space, namely extremely low temperatures and pressures. In particular, identifying radiative association channels experimentally is far from straightforward\cite{Ascenzi2007,Ascenzi2010,Aysina2013,Rap2022,Rap2024}, and this is precisely where ICR setups such as the one used here can provide definitive answers.

\section*{Acknowledgments}
This work was supported by the Agence Nationale pour la Recherche, ANR grant No ANR-21-CE30- 0010, SynPAHcool. The running costs for cryogenic fluids, required for the superconducting magnet of the PIRENEA setup, were also supported by recurrent funding for the Nanograin platform at IRAP, as well as by the Thematic Action “Physique et Chimie du Milieu Interstellaire” (PCMI) of the INSU Programme National “Astro”, with contributions from CNRS Physique, CNRS Chimie, CEA, and CNES. The authors gratefully acknowledge Loïc Noguès, David Murat and Odile Coeur-Joly for their sustained technical support of the PIRENEA setup over many years. 
AS thanks the computing mesocenter CALMIP (“CALcul en MIdi Pyr\'en\'ees”, UAR~3667 of CNRS) for generous allocation of computer resources (project p17002).
Finally, the authors sincerely thank the reviewers for their valuable suggestions, which significantly improved the manuscript -- particularly by guiding the development of a plausible chemical scenario.

\section*{Associated content}

\subsection*{Data Availability Statement}
The dataset associated with this work can be found under 10.5281/zenodo.19106840

\section*{Safety Statement}
No unexpected or unusually high safety hazards were encountered during the course of this study. All experiments were conducted in accordance with standard laboratory safety 

\section*{Author Contributions}
A. I. L.: Investigation, Formal analysis, Data curation, Visualization, Writing -- original draft, Writing -- review \& editing. \\
A. B.: Instrumentation, Methodology, Investigation, Validation, Resources. \\
A. S.: Methodology, Investigation, Computation, Resources, Writing -- original draft, Writing -- review \& editing. \\
C. J.: Conceptualization, Methodology, Investigation, Formal analysis, Resources, Writing -- original draft, Writing -- review \& editing, Supervision, Project administration, Funding acquisition.\\
All authors have read and agreed to the published version of the manuscript.

\break

\bibliography{references}

\appendix


\section*{Supplementary material}

# Supplementary material for: Acenaphthene Derivatives as Signatures of $C_{11}H_9^+$ Reactivity with Methylated Naphthalenes

Ana I. Lozano,[†] Anthony Bonnamy,[†] Aude Simon,[‡] and Christine Joblin[*,†]

†*Institut de Recherche en Astrophysique et Planétologie (IRAP), Université de Toulouse, CNRS, CNES, 9 Avenue du Colonel Roche, 31028 Toulouse, France*

‡*Laboratoire de Chimie et Physique Quantiques LCPQ/FeRMI, Université de Toulouse, CNRS, 118 Route de Narbonne, 31062 Toulouse, France*

E-mail: christine.joblin@cnrs.fr

**Number of pages:** 15

**Number of figures:** 10

**Number of tables:** 6

- Figure S1: Mass spectra illustrating the production of $C_{11}H_9^+$ from 1,4-diMeNp.
- Figure S2: Mass spectra for reactions between $C_{11}H_9^+$ and diMeNp.
- Figure S3: Mass spectra showing the protofragments of $C_{13}H_{13}^+$ and $C_{12}H_{11}^+$ ions.
- Figure S4: Photofragmentation kinetic curves of $C_{13}H_{13}^+$.
- Figure S5: Photofragmentation kinetic curves of $C_{11}H_9^+$ fitted using a single exponential decay model.
- Table S1: Fitting parameters of the experimental photofragmentation kinetic curves.
- Table S2: Branching ratios for the dissociation of 1,4-, 1,5-, and 2,3-diMeNp cations, obtained from MD/DFTB simulations over a 1 ns timesale at two different internal energies.
- Figure S6: Snapshot for the dissociation of 2,3-diMeNp$^+$ resulting in 1-Np$CH_2^+$.
- Figure S7: MPD spectrum of $C_{13}H_{13}^+$ species and their photofragments.
- Figure S8: Branching ratio for the two dissociation channels of $C_{13}H_{13}^+$ species, derived from the MPD spectrum shown in Figure S7
- Table S3: Vertical electronic excitations energies of acenaphthene radical cation.
- Table S4: H-loss energies from the different isomers of acenaphthylene cation.
- Figure S9: Molecular structure of the two $C_{12}H_{11}^+$ isomers bearing a $CHCH_3$ chain.
- Table S5: Branching ratios for the isomerization of 1-EtNp$^+$ and 2-EtNp$^+$ (Figure S9).
- Table S6: Relative enthalpies at 0 and 298 K for the formation of the different isomers of $C_{13}H_{13}^+$ and naphthalene, starting from 2,3-diMeNp and 2-Np$CH_2^+$.
- Figure S10: Geometries of the isomers of $C_{13}H_{13}^+$ considered in Table S6.

# S1 Mass Spectra

## S1.1 $m/z$ 141 formation from 1,4-dimethylnaphthalene

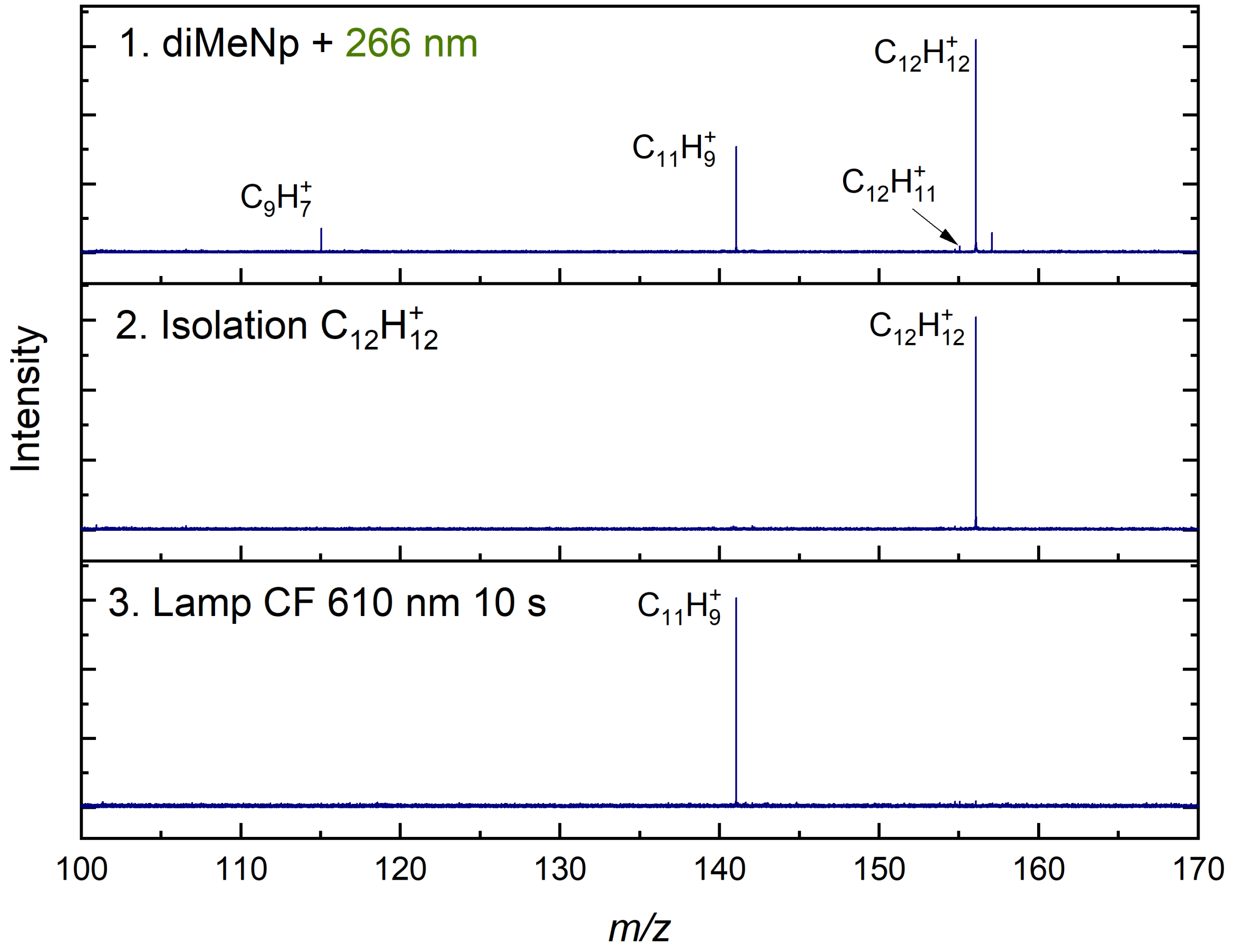


Figure S1: Mass spectra illustrating the procedure used to produce the $C_{11}H_9^+$ from 1,4-diMeNp. The steps include: 1- irradiation of the neutral precursor with the 266 nm laser during 2 s, 2- isolation of $C_{12}H_{12}^+$, and 3- irradiation with the Xe lamp (CF at 610 nm) during 10 s.

## S1.2 Ion-molecule reaction experiments

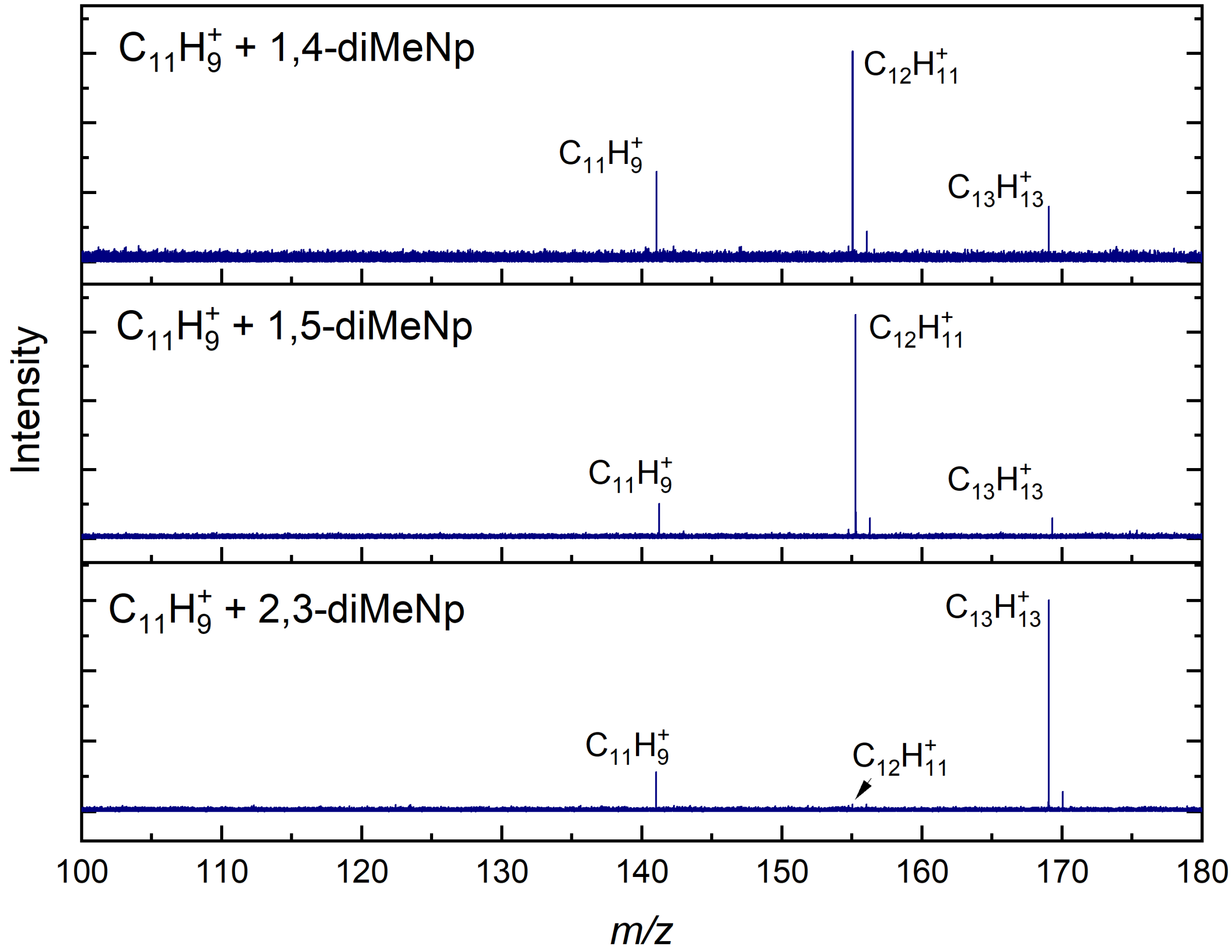


Figure S2: Mass spectra for reactions between $C_{11}H_9^+$ and different isomers of diMeNp. The precision of the ICR mass spectrometer allows us to confirm the identification of the two reaction products with derived masses at $m/z$ 155.084 and 169.099.

### S1.3 Photodissociation $C_{12}H_{11}^+$ and $C_{13}H_{13}^+$

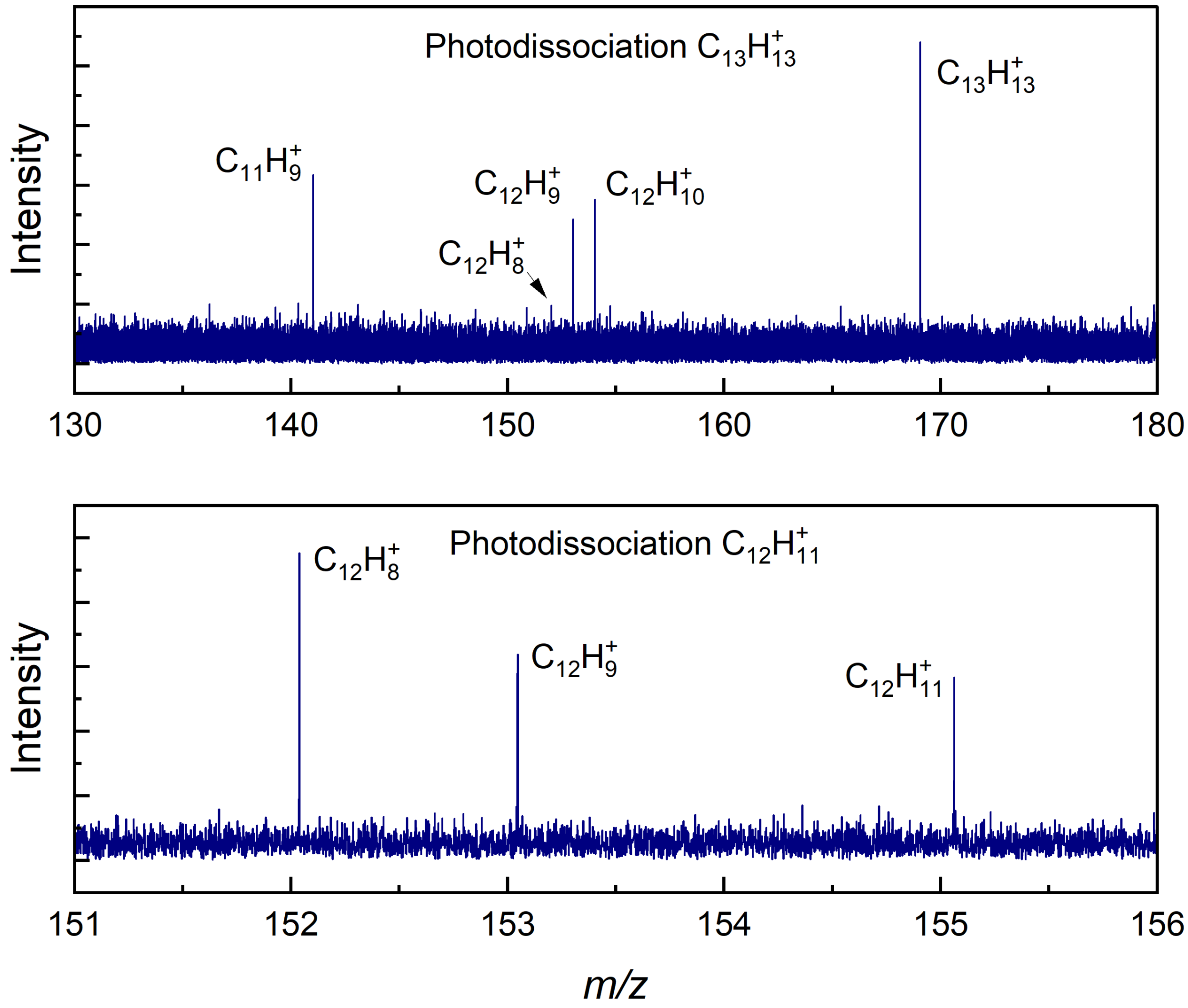


Figure S3: Mass spectra illustrating the photofragments of $C_{13}H_{13}^+$ (top) and $C_{12}H_{11}^+$ (bottom) under broad-band excitation with the Xe lamp (CF at 400 nm).

## S2 Photoprocessing Kinetic Curves

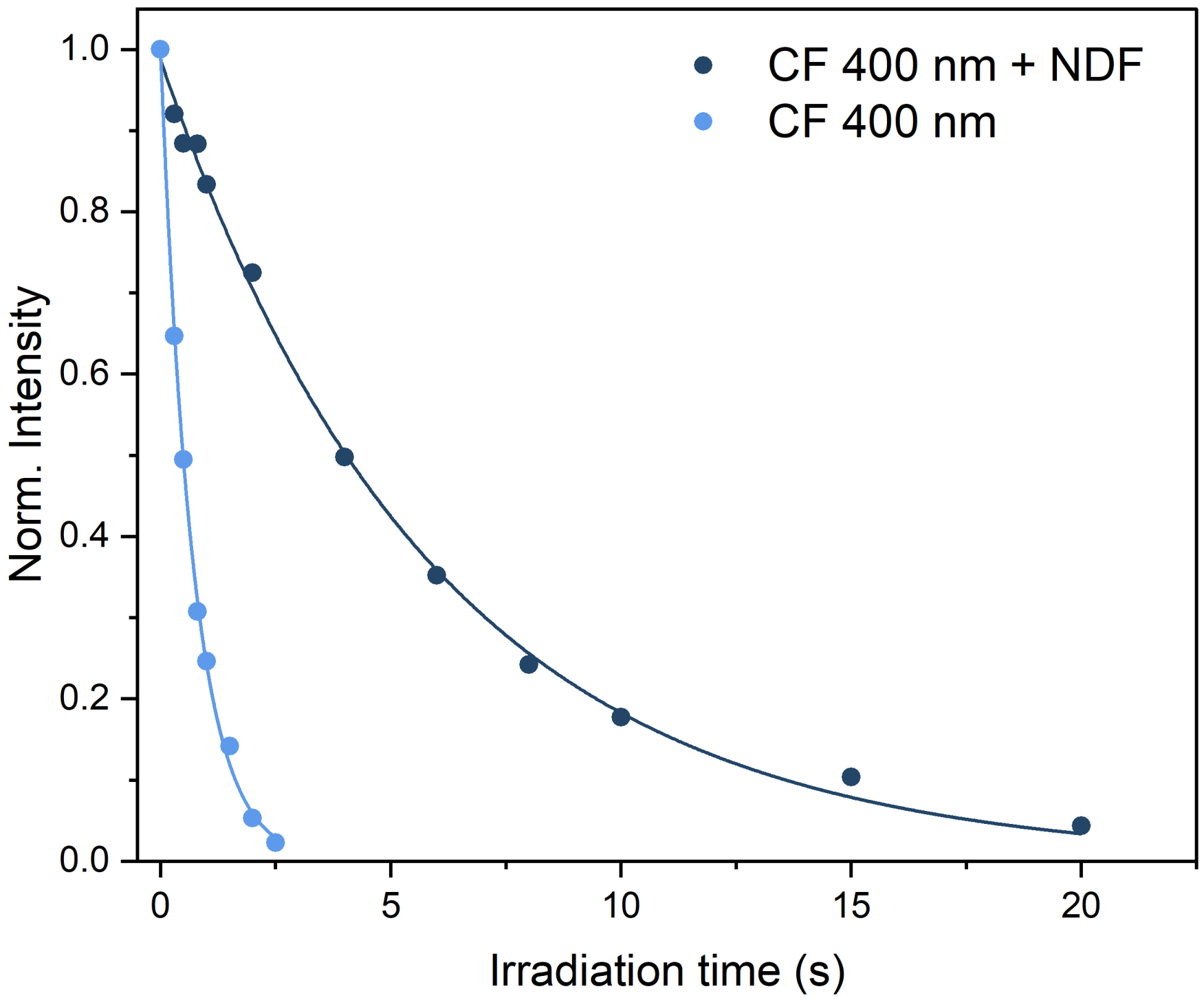


Figure S4: Normalized intensity of $C_{13}H_{13}^{+}$ species as a function of the irradiation time with the Xe arc lamp using a color filter (CF) at 400 nm and with or without and additional neutral density filter (NDF) of 37% transmission. The solid lines represent fits based on an exponential decay model (Table S1).

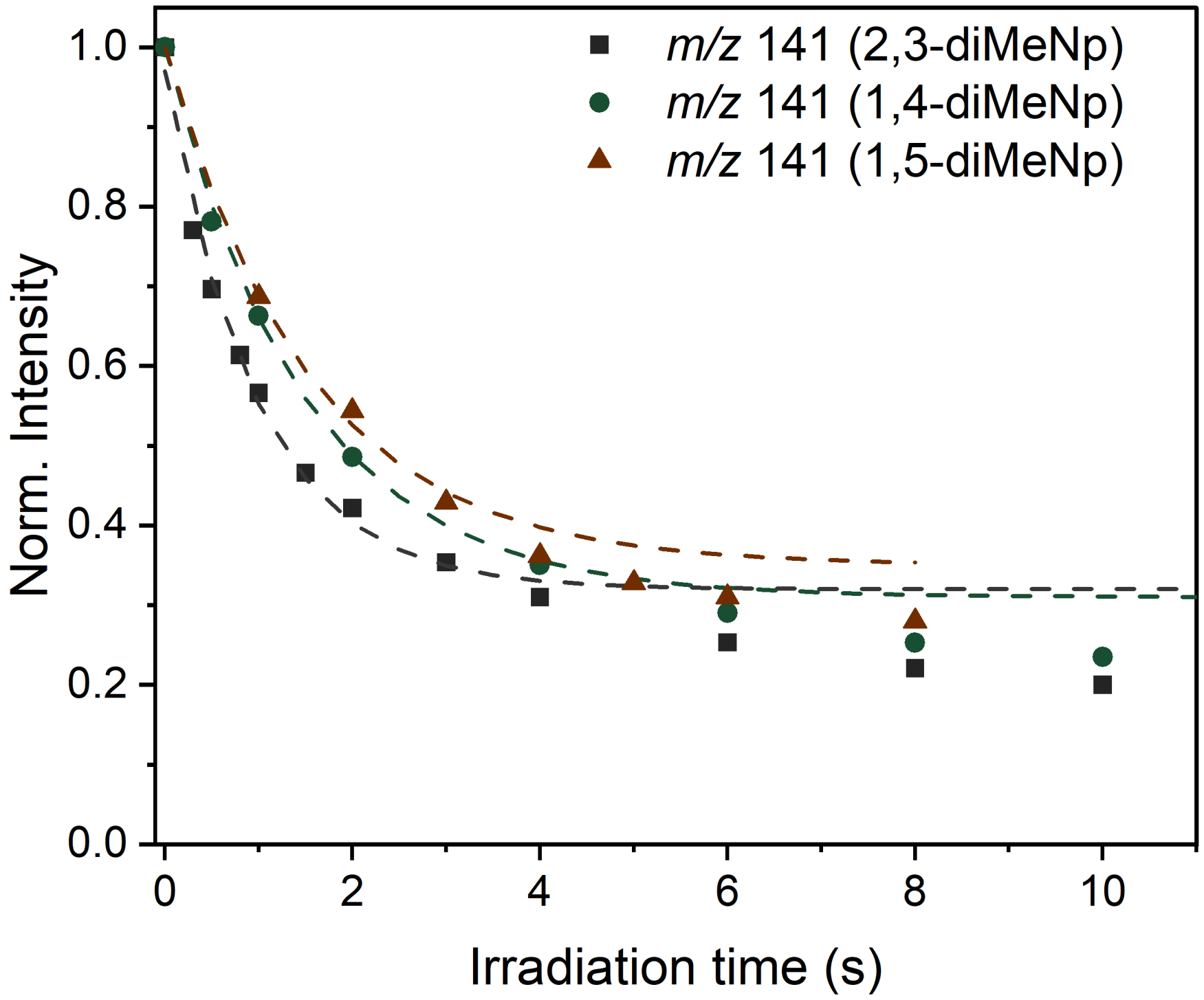


Figure S5: Photofragmentation kinetic curves of $C_{11}H_9^+$ ions (produced from different precursors) under irradiation from the Xe arc lamp with a CF at 400 nm. The data are the same as in Figure **??**, but the experimental points were fitted using a single exponential decay model ($A_0 + Ae^{-kt}$) obtained by considering only the experimental points below 4 s (dashed lines).

Table S1: $k$ values derived from the kinetic curves shown in Figures **??**, **??**, and S4, using either a single exponential or a two exponential decay model. In the latter case, relative populations (A) associated with each $k$ value are given in parentheses.

| Species | Precursor | CF 400 nm + NDF (37%) $k$ ($s^{-1}$) | CF 400 nm $k$ ($s^{-1}$) ($A$) |
|---|---|---|---|
| $C_{11}H_9^+$ ($m/z$ 141) | 2,3-diMeNp | | 0.86 (70 %)<br>0.0394 (30 %) |
| | 1,4-diMeNp | | 0.721 (65 %)<br>0.0358 (35 %) |
| | 1,5-diMeNp | | 0.654 (63 %)<br>0.0342 (37 %) |
| $C_{12}H_{11}^+$ ($m/z$ 155) | 1-MeNp | 0.134 | 0.992 |
| | 1,4-diMeNp | 0.105 | 0.741 |
| | 1,5-diMeNp | 0.096 | 0.75 |
| $C_{13}H_{13}^+$ ($m/z$ 169) | 2,3-diMeNp | 0.167 | 1.39 |

# S3 MD simulations for dimethylnaphthalene cations

Table S2: Branching ratios for the dissociation of 1,4-, 1,5-, and 2,3-diMeNp cations, obtained from MD/DFTB simulations over a 1 ns timescale at two different internal energies. The values were averaged over 180 simulations with randomly assigned velocity distributions. Note: # denotes the number of occurrences.

| Internal energies (eV) | 15.0 (1 ns) | | | 16.5 (1 ns) | | |
|---|---|---|---|---|---|---|
| Precursor | 1,4-diMeNp | 1,5-diMeNp | 2,3-diMeNp | 1,4-diMeNp | 1,5-diMeNp | 2,3-diMeNp |
| #1-$NpCH_2^+$/#$C_{11}H_9^+$ | 0.50 | 0.85 | 0.13 | 0.62 | 0.73 | 0.17 |
| #2-$NpCH_2^+$/#$C_{11}H_9^+$ | 0.13 | 0.03 | 60 | 0.09 | 0.08 | 0.50 |
| #$BzTr^+$/#$C_{11}H_9^+$ | 0.27 | 0.03 | 0.20 | 0.13 | 0.03 | 0.05 |
| #Others/#$C_{11}H_9^+$ | 0.10 | 0.09 | 0.07 | 0.16 | 0.15 | 0.26 |

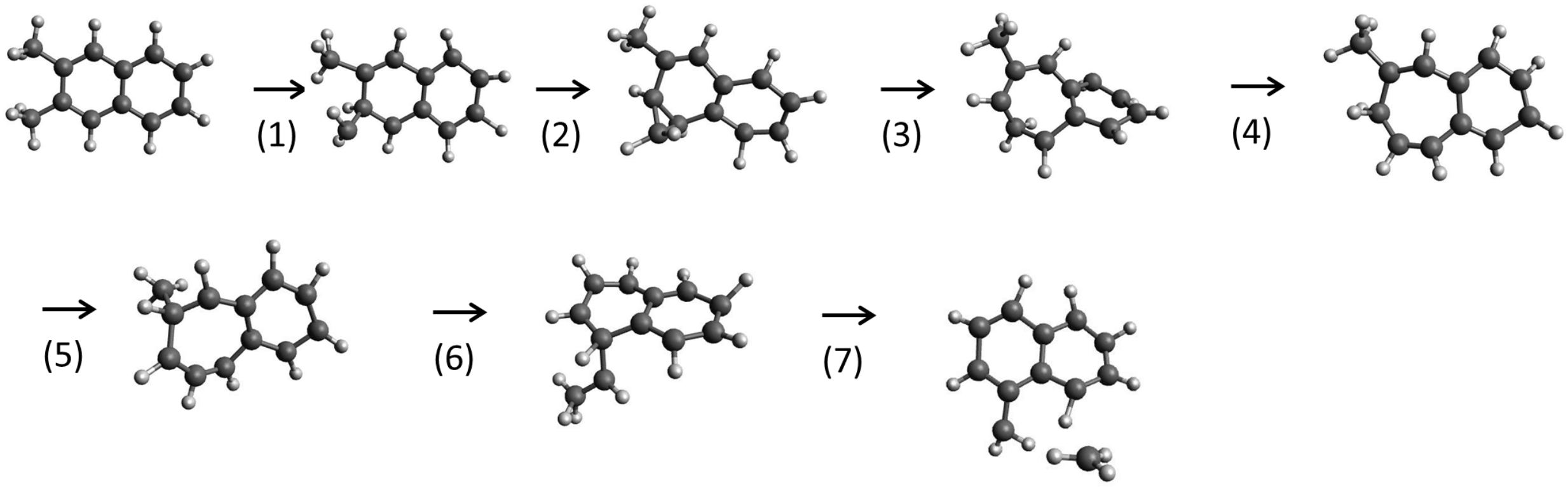


Figure S6: Snapshots from a 15.0 eV MD/DFTB simulation illustrating the dissociation of 2,3-diMeNp$^{\bullet+}$ into 1-$NpCH_2^+$.

The following elementary steps are observed : (1) H migration, (2) $CH_2$ migration to form two rings [6+3], (3) Formation of a 7-member ring, (4) and (5): H migration, (6) Ring shrinking [6+3] (reverse of (3)) and opening: this step can lead to both isomers (1- or 2-$NpCH_2^+$), (7) $CH_3$ loss.

## S4 Spectroscopy

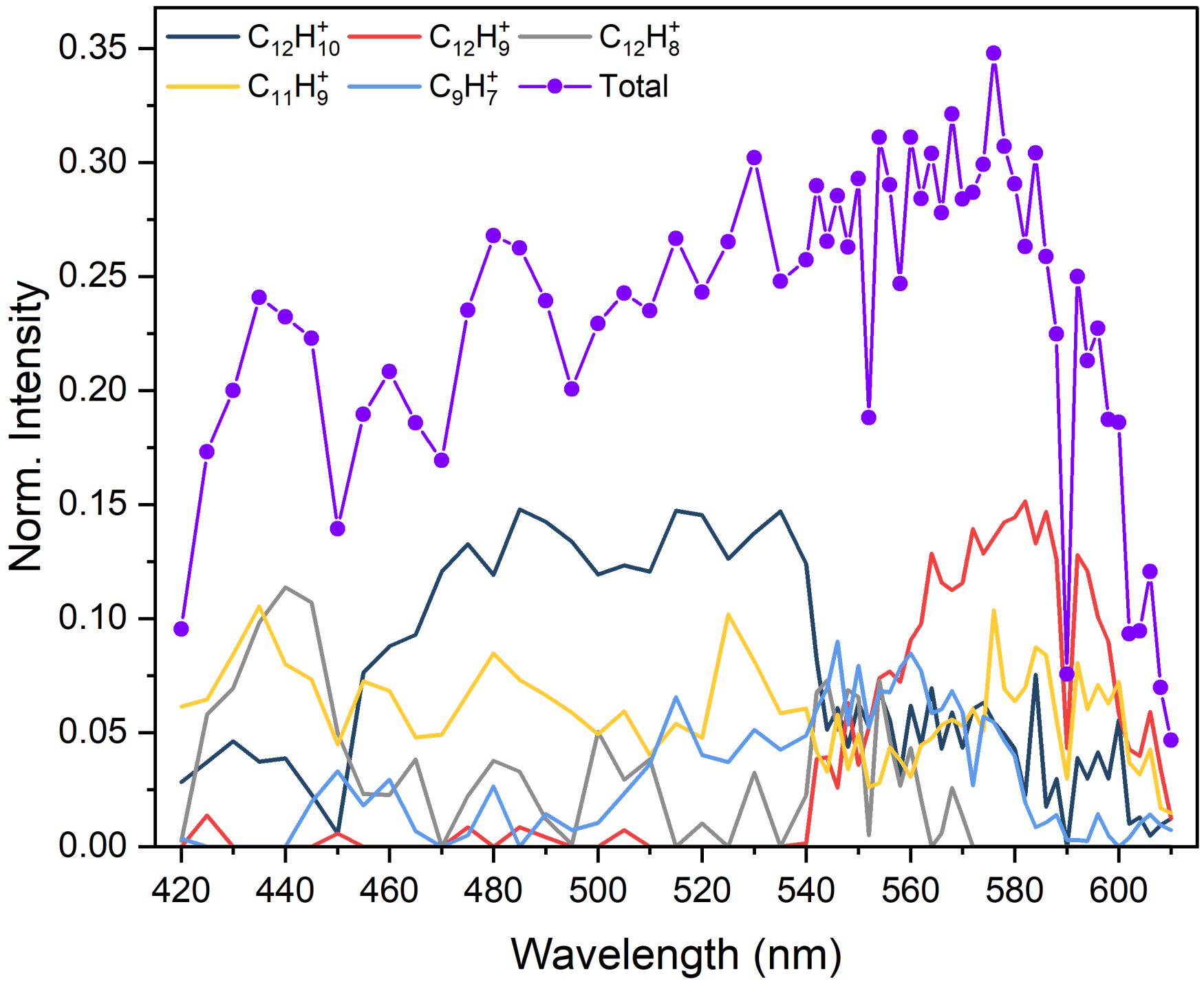


Figure S7: MPD spectrum of $C_{13}H_{13}^+$ species and their photofragments after 10 s of irradiation with the OPO laser at 5.5 mJ. The spectrum corresponds to a single scan; thus, only global variations in intensity can be analyzed with confidence.

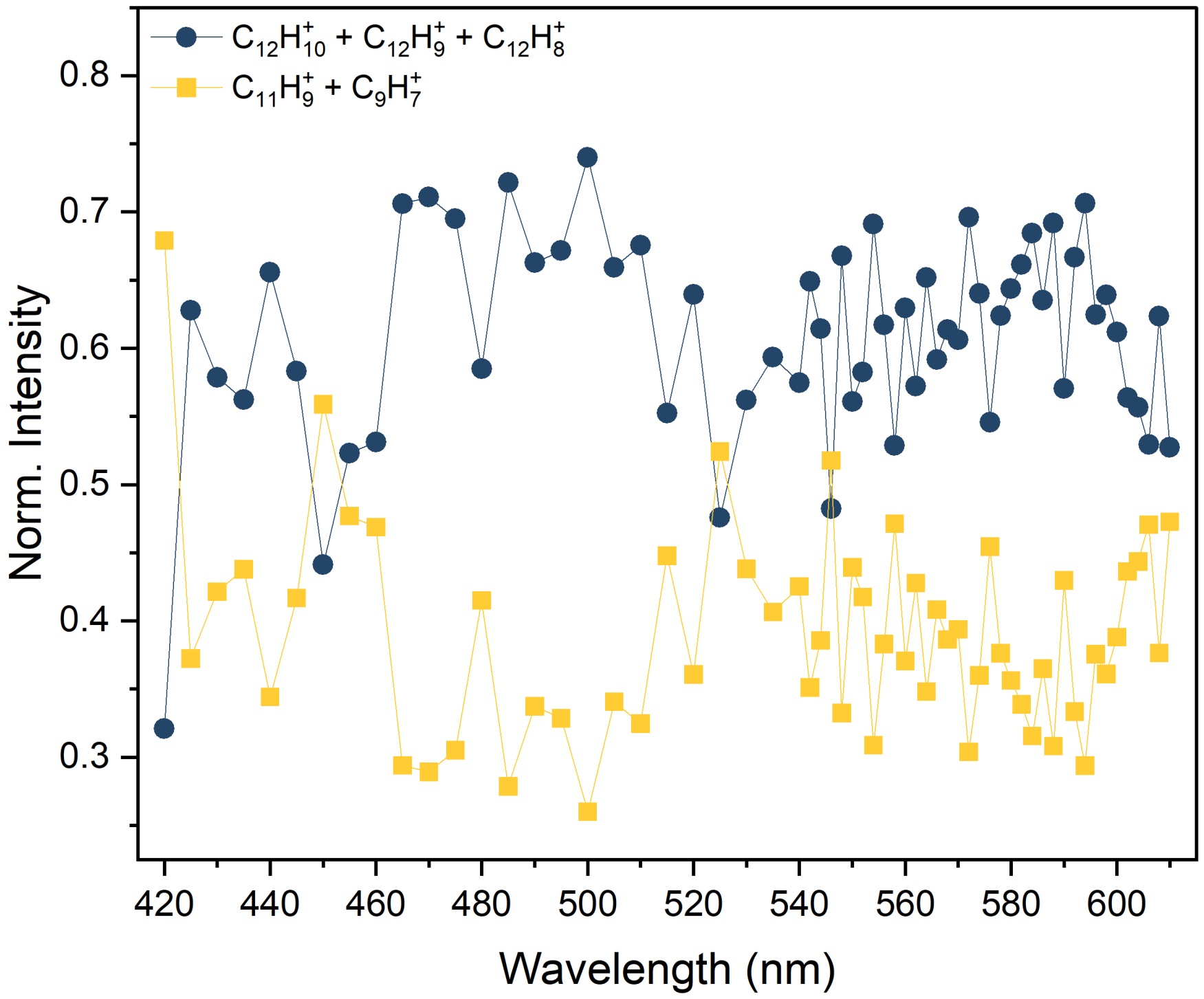


Figure S8: Branching ratio for the two dissociation channels of $C_{13}H_{13}^{+}$ species, derived from the MPD spectrum shown in Figure S7. Average values of 0.609±0.075 and 0.391±0.075 are obtained for the $C_{12}H_{10}^{+}$ and $C_{11}H_{9}^{+}$ channels, respectively.

Table S3: Vertical electronic excitations of acenaphthene radical cation: symmetries, energies with respect to that of the ground state ($\Delta E$) and oscillator strengths ($f$) corresponding to the first three electronic excited states of acenaphthene radical cation. The time-dependent DFT calculations were performed at the PBE0/cc-pVTZ level using the B3LYP/6-31G(d,p) optimized geometry of the ground electronic state.

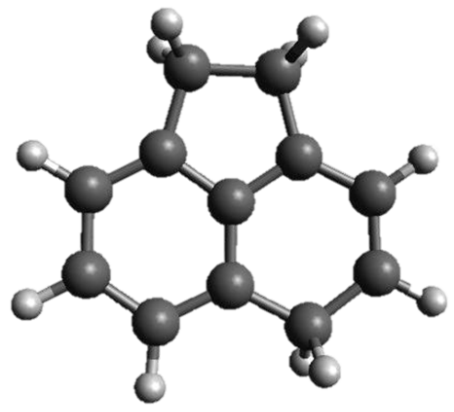

ACN$^{\bullet+}$

*m/z* 154

| Excited state | $\Delta E$ (eV) | $\lambda$ (nm) | $f$ |
|---|---|---|---|
| $1\,^2B_2$ | 1.41 | 877 | 0 |
| $2\,^2B_2$ | 2.13 | 582 | 0.0488 |
| $1\,^2A_1$ | 3.03 | 410 | 0.0044 |

# S5 H-loss Energy Calculations

Table S4: H-loss energies (in eV) from the different isomers of protonated acenaphthylene (ACYN+$H^+$) determined at the B3LYP/6-31G(d,p) (DFT column) or CCSD(T)/cc-pvtz//B3LYP/6-31G(d,p) (CCSD(T) column) levels of theory. Zero-point energies computed at the B3LYP/6-31G(d,p) level of theory were taken into account. The yellow labels indicate the position of the proton.

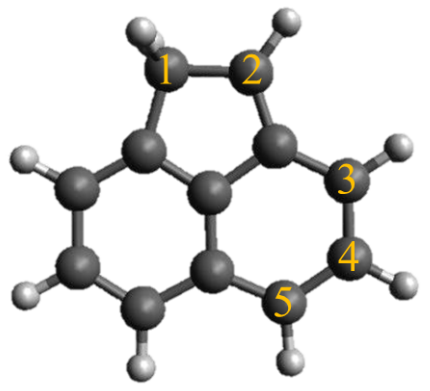


[ACYN + $H^+$]

*m/z* 153

| H Position | DFT | CCSD(T) |
|---|---|---|
| 1/2 | 3.21 | 3.30 |
| 3 | 2.43 | 2.58 |
| 4 | 2.49 | 2.65 |
| 5 | 2.70 | 2.82 |

# S6 MD simulations for $C_{12}H_{11}^+$

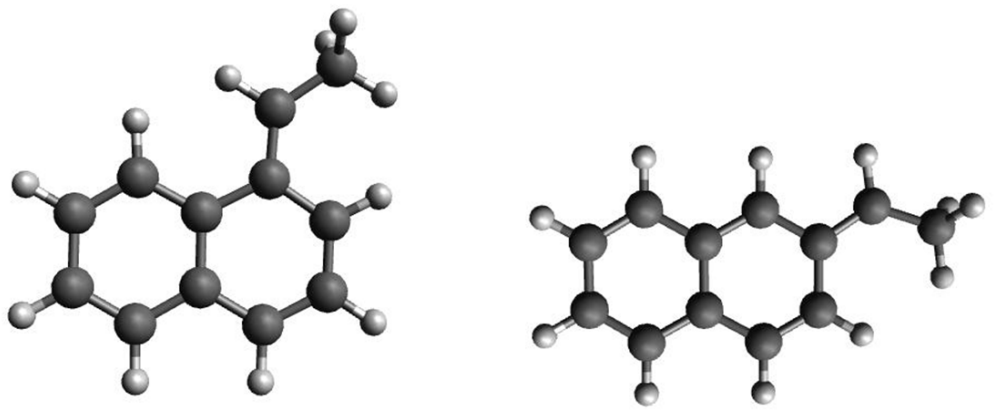

1-EtNp$^+$ *m/z* 155

2-EtNp$^+$ *m/z* 155

Figure S9: Molecular structures of the two $C_{12}H_{11}^+$ isomers bearing a $=CH-CH_3$ chain at positions 1 (1-EtNp$^+$) and 2 (2-EtNp$^+$) considered for the MD simulations.

Table S5: Branching ratios for the isomerization of 1-EtNp$^+$ and 2-EtNp$^+$ (Figure S9), obtained from MD/DFTB simulations over a 100 and 500 ps timescale at 13 eV internal energy. The values were averaged over 180 simulations with randomly assigned velocity distributions. [ACN + H$^+$] refers to protonated acenaphthene, where the position of the extra H is indicated by the accompanying number.

| Time | 100 ps | | 500 ps | |
|---|---|---|---|---|
| Parent | 1-EtNp$^+$ | 2-EtNp$^+$ | 1-EtNp$^+$ | 2-EtNp$^+$ |
| 1-EtNp$^+$ | 0.40 | 0 | 0.15 | 0.03 |
| 2-EtNp$^+$ | 0.02 | 0.34 | 0.08 | 0.23 |
| [ACN+H$^+$] position 5 | 0.09 | 0 | 0.09 | 0.01 |
| [ACN+H$^+$] position 3 | 0.02 | 0 | 0.09 | 0.02 |
| [ACN+H$^+$] position 4 | 0.02 | 0 | 0.02 | 0 |
| Others | 0.48 | 0.66 | 0.58 | 0.73 |

# S7 Computational details and additional results for Section 4.2 "Formation of acenaphtene derivatives: building chemical scenarii"

**Computational details.** The relative enthalpies at 0 K and 298 K of the products and possible RIs represented in Figure **??**, relative to the reactants, were computed at the B3LYP-GD3/aug-cc-pvtz//B3LYP-GD3/6-31G(d,p) level of theory within the Gaussian16 package. Except in specific cases explained hereafter, only local B3LYP-GD3/6-31G(d,p) optimizations were performed. For the calculations performed in this section, dispersion interactions were taken into account taken into account through the empirical term introduced by Grimme and collaborators[1] due to hypothesis of the formation of the initial molecular complex RI 1 and RI 1'. In the case of the ion-molecule complexes RI 1 and RI 1', the geometries were obtained through a three-step procedure : (i) the potential energy surface (PES) was explored using a parallel tempering Monte-Carlo algorithm (PTMC) with the electronic structure computed at the SCC-DFTB level - using the hamiltonian described in Rapacioli et al.[2], constraining the positive charge to be on the 1-NapCH$_2$ monomer. The PTMC simulations were run for 1,000,000 steps with temperatures ranging from 40 K to

400 K ; exchanges between all temperatures were allowed and the moveclass was restricted to intermolecular motions. A repulsive wall (sphere of 13 Å radius centered on the center of mass of the system) was added to prevent from evaporation. (ii) The 100 most stable isomers were retrieved and locally optimized at the same charge constrained SCC-DFTB level with all atoms optimization. (iii) The most stable isomers were then optimized at the DFT (B3LYP-GD3/6-31G(d,p)) level. The derived geometries for RI 1 and RI 1' are reported in Figure **??**. For RI 2 and RI 2', the geometries were first locally optimized at the SCC-DFTB level (with no charge constraint) and subsequently optimized at the DFT level.

Table S6: Relative enthalpies at 0 and 298 K (in eV) for the formation of the different isomers of $C_{13}H_{13}^+$ and naphthalene, starting from 2,3-diMeNp and 2-$NpCH_2^+$, computed at the B3LYP-GD3/aug-cc-pvtz//B3LYP-GD3/6-31G(d,p) level of theory. The optimized geometries are reported in Figure S10. The following abbreviations were used: Phen for phenalene, Acyn for acenapthylene, Nap for naphthalene, and Acen for acenaphthene.

| $C_{13}H_{13}$ Isomer | ΔH(0K) | ΔH(298K) |
|---|---|---|
| 2,3,6-trihydro-1H-Phen$^+$ | -0.61 | -0.65 |
| 2,5-dihydro-1-methylAcyn$^+$ | -0.50 | -0.52 |
| 2-isopropylNap$^+$ | -0.52 | -0.52 |
| 5-methylAcen$^+$ | -0.46 | -0.49 |
| 5H-cyclopenta[b]Nap$^+$ | -0.31 | -0.34 |
| 1H-cyclopenta[a]Nap$^+$ | -0.24 | -0.28 |
| 2-propylNap$^+$ | -0.23 | -0.24 |
| 5H-cyclopenta[a]Nap$^+$ | -0.19 | -0.22 |
| 2H-cyclopenta[b]Nap$^+$ | 0.03 | -0.01 |

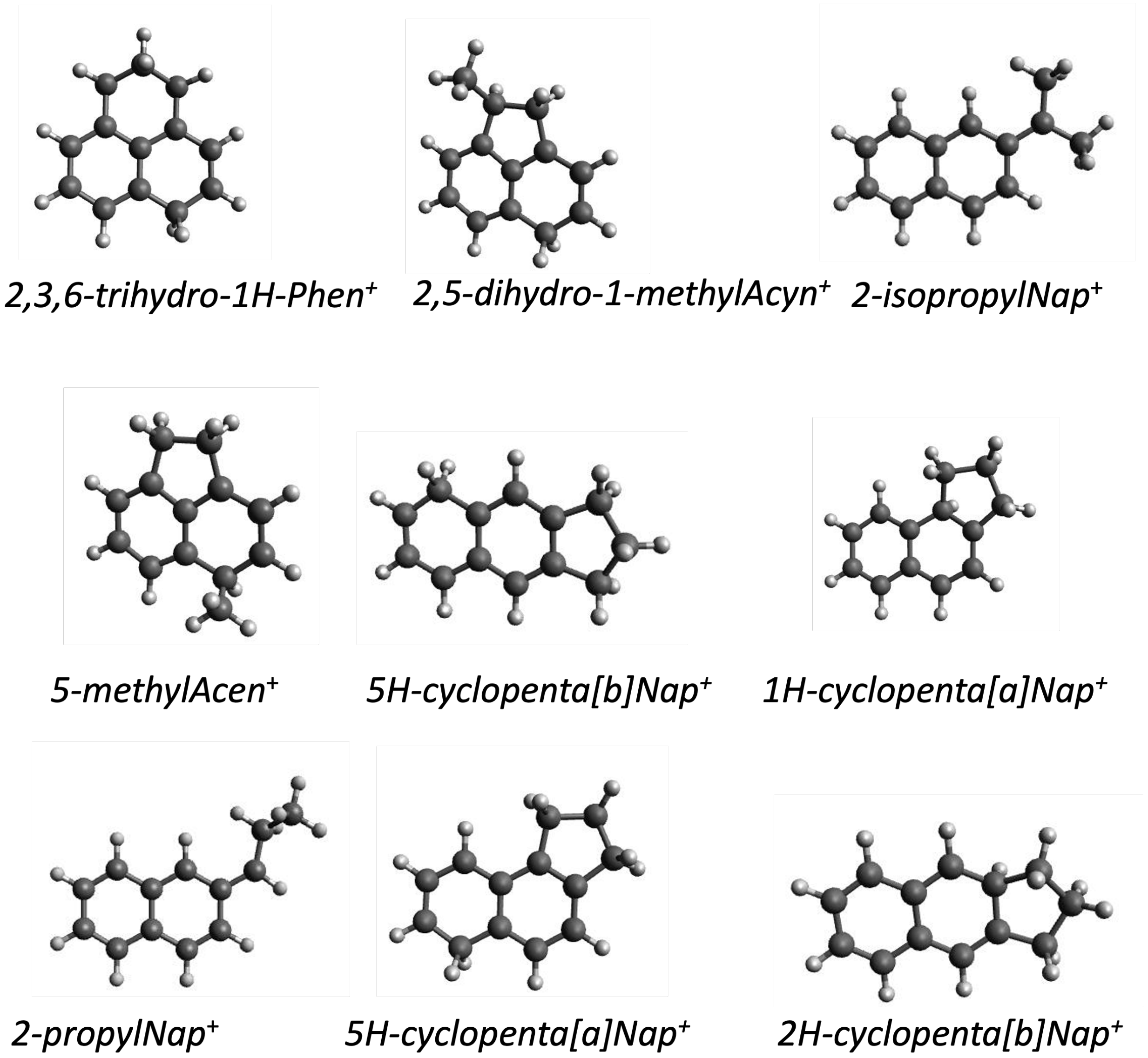


Figure S10: Geometries of the isomers of $C_{13}H_{13}^{+}$ considered in Table S6, reported from the most stable one to the least stable one from top left to bottom right. The following abbreviations were used: Phen for phenalene, Acyn for acenapthylene, Nap for naphthalene, and Acen for acenaphthene.



\end{document}